\documentclass[%
 reprint,
 amsmath,amssymb,
 aps,
 pra,
]{revtex4-1}

\usepackage{graphicx}
\usepackage{dcolumn}
\usepackage{bm}
\usepackage{enumitem}
\usepackage{placeins}
\usepackage{hyperref}
\usepackage{mathtools}

\DeclarePairedDelimiter\floor{\lfloor}{\rfloor}
\newcommand{\dif}[2]{\frac{\text{d} #1}{\text{d} #2}}
\newcommand{\diff}[2]{\frac{\partial #1}{\partial #2}}
\newcommand{\vardiff}[2]{\frac{\delta #1}{\delta #2}}
\DeclareMathOperator*{\argmin}{arg\,min}

\begin{document}


\title{Quantum Optimal Control in a Chopped Basis: Applications in Control of Bose-Einstein Condensates}

\author{J. J. W. H. S\o rensen}
\author{M. O. Aranburu}
\author{T. Heinzel}
\author{J. F. Sherson}
\affiliation{Department of Physics and Astronomy, Aarhus University, Ny Munkegade 120, 8000 Aarhus C, Denmark}
\email{sherson@phys.au.dk}
\date{\today}

\begin{abstract}
We discuss quantum optimal control of Bose-Einstein condensates trapped in magnetic microtraps. The objective is to transfer a condensate from the ground state to the first-excited state. This type of control problem is typically solved using derivative-based methods in a high-dimensional control space such as gradient-ascent pulse engineering (\textsc{grape}) and Krotov's method or derivative-free methods in a reduced control space such as Nelder-Mead with a chopped random basis (\textsc{crab}). We discuss how these methods can be combined in gradient optimization using parametrization (\textsc{group}) including the finite bandwidth of the control electronics. We compare these methods and find that \textsc{group} converges much faster than Nelder-Mead with \textsc{crab} and achieves better results than \textsc{grape} and Krotov's method on the control problem presented here.
\end{abstract}

\maketitle


\section{Introduction}
Technological advances in the experimental toolbox in physical chemistry and atomic, molecular, and optical physics currently enable exciting new developments in the manipulation of complex quantum systems. Gradually, the focus is shifting from verifying the validity of theoretical models towards controlling and manipulating quantum systems for specific technological applications \cite{bloch2008many,gordon1997active}. Some examples of this trend are quantum state preparation \cite{sayrin2011real,bucker2011twin}, atomic clocks \cite{bloom2014optical}, quantum based sensors \cite{lucke2011twin,riedel2012atom,gross2012nonlinear}, quantum simulators \cite{bloch2012quantum}, and quantum computers \cite{anderlini2007controlled,kielpinski2002architecture}. 

These applications require the ability to steer the quantum dynamics precisely using external control fields. The quality of a control is measured by a cost function, which can describe e.g. distance to some target state \cite{hohenester2007optimal}, similarity with a unitary gate operator \cite{palao2002quantum}, or the amount of experimental signal \cite{kaiser2004optimal}. Quantum optimal control (QOC) is a framework that enables the design of control strategies that achieve the desired dynamics \cite{glaser2015training,werschnik2007quantum,peirce1988optimal}. QOC has been studied in a wide range of physical systems \cite{hohenester2007optimal,goerz2017charting,koch2004stabilization,nobauer2015smooth,caneva2011chopped}. Central in QOC are local optimization algorithms that maximize or minimize the cost function. These local algorithms can broadly be divided along two axes \cite{globalPaper}. The first axis is derivative-based versus derivative-free algorithms and the second is optimization in a full or reduced-basis for admissible controls. It is also possible to extend the local algorithms with global optimization \cite{globalPaper}.

Derivative-free algorithms operate by directly evaluating the cost functional in carefully selected points. A particularly prevalent example in QOC is the Nelder-Mead algorithm \cite{caneva2011chopped,doria2011optimal,van2016optimal,brouzos2015quantum}. More recently, it has been proposed to use gradient-free methods in QOC such as Brent's principal axis \cite{goetz2016maximizing}. On the other hand, derivative-based methods use both the functional and derivative information, which can speed up the convergence rate. Important examples from QOC are the gradient-ascent pulse engineering (\textsc{grape}) \cite{khaneja2005optimal} and Krotov's method \cite{sklarz2002loading,reich2012monotonically}.
\begin{figure}
\includegraphics[width=\columnwidth]{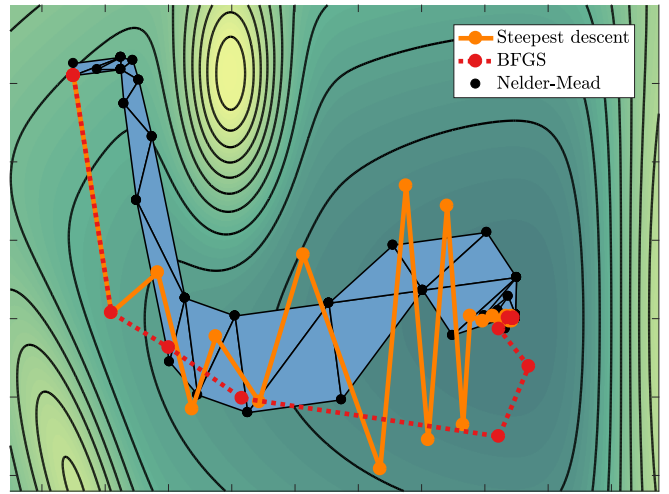}
\caption{(Color online) Comparison of gradient-based and gradient-free optimization methods in an artificial landscape. The shaded blue triangles show the gradient-free method, Nelder-Mead. The solid orange line and the dotted red line are gradient-based algorithms steepest descent and \textsc{bfgs} respectively. Steepest descent exhibits the characteristic zigzag type behavior, which \textsc{bfgs} avoids due to the inverse Hessian approximation. Nelder-Mead, steepest descent, and \textsc{bfgs} respectively use 45, 34, and 15 iterations for convergence.}
\label{fig:cartoon}
\end{figure}

In a reduced-basis method, the control space is parametrized by a set of smooth functions \cite{caneva2011chopped}. This is motivated by the fact that the high-quality solutions often lie in some lower dimensional subspace of the full control space \cite{lloyd2014information}, so a proper low dimensional parametrization of the control space can \textit{a priori} steer the optimization in the correct direction. A prominent example is chopped random basis (\textsc{crab}) \cite{caneva2011chopped,doria2011optimal}. The introduction of a random chopped basis is on the idea that the randomness allows one to explore a range of different bases, which can lead to improved results. A too-low-dimensional parametrization can introduce artificial traps since the parametrization is no longer able to adequately describe the optimal solutions \cite{rach2015dressing}. Within QOC, reduced-basis methods are typically only used in combination with derivative-free algorithms such as Nelder-Mead \cite{caneva2011chopped,doria2011optimal,brouzos2015quantum,van2016optimal,kelly2014optimal}.

Ideally, one should combine both of these two approaches and perform derivative-based optimization in a reduced-basis. In the past years, steps have been taken to implement this in a number of different ways \cite{machnes2015gradient,von2008computational,lucarelli2016quantum,skinner2010optimal,motzoi2011optimal}. We will collectively refer to such methods as gradient optimization using parametrization (\textsc{group}). In part due to the frequent success
of derivative-free search in a reduced-basis, \textsc{group}-type
methods have not been widely adopted in the QOC community. However, in recent years, there has been a growing awareness that the search for time-optimal solutions in quantum engineering leads to an exponentially growing computational complexity \cite{sorensen2016exploring,globalPaper,bukov2017reinforcement,day2018glassy,bukov2018broken}, which will necessitate algorithmic improvements. In this work, we present the \textsc{group} formalism and show how to combine \textsc{grape} and \textsc{crab} type methods. In addition, we clearly demonstrate the efficiency of the \textsc{group}-methodology to the field of quantum engineering by applying it to a high profile challenge in the field and presenting a detailed comparison of performance with conventional algorithms. In this work we also demonstrate how the \textsc{group} formalism can include a filter function on the controls.

This paper discusses quantum control of Bose-Einstein condensates manipulated in a magnetic microtrap. These systems are particularly challenging to control due to the nonlinearity in the equations of motion.  We will investigate fast excitation from the ground state to the first-excited state in a single well. This system has attracted much attention in QOC and it has been investigated both experimentally and theoretically \cite{bucker2013vibrational,van2016optimal,jager2014optimal,hohenester2007optimal}. Due to the interest in this system, we see it as an ideal test bed for quantum control algorithms.

Here, we present a direct comparison of derivative-based search in a reduced-basis (\textsc{group}) with the other state-of-the-art algorithms Krotov's method, Nelder-Mead using \textsc{crab}, and \textsc{grape}. We find that \textsc{group} not only converges faster, but also achieves better end fidelity. This work supplements and extends previous efforts in comparing QOC algorithms \cite{jager2014optimal,machnes2011comparing}.

Derivative-based methods generally converge faster than derivative-free \cite{machnes2011comparing,nocedal2006numerical}. Derivative-based methods typically employ either steepest descent or the quasi-Newton method of Broyden-Fletcher-Goldfarb-Shanno (\textsc{bfgs}) \cite{nocedal2006numerical}. \textsc{Bfgs} is typically faster than steepest descent since it avoids steepest descent's characteristic inefficient zigzag motion close to the optimum \cite{machnes2011comparing,nocedal2006numerical}. \textsc{Bfgs} achieves this by gradually building an approximation of the Hessian using the past gradients \cite{nocedal2006numerical}. In this paper we only use \textsc{bfgs} type descent with \textsc{grape} and \textsc{group}. A graphical illustration of the difference in convergence behavior between a derivative free Nelder-Mead and the two derivative-based methods is given in Fig. \ref{fig:cartoon}.

This paper is organized as follows. In Sec. \ref{sec:TheControlProblem}, we introduce the control problem of transferring a condensate from the ground state to the first-excited state. In Sec. \ref{sec:QuantumOptimalControl}, we present the \textsc{group} methodology and the other different quantum control algorithms. In Sec. \ref{sec:filter}, we present how to include the finite bandwidth of the control electronics. In Sec. \ref{sec:Results}, we compare the different algorithms. Finally, we conclude the paper in section \ref{sec:Conclusion}.

\section{The Control Problem} \label{sec:TheControlProblem}
We will discuss the manipulation of a condensate trapped on an atom chip \cite{bucker2011twin}. In the experiment described in Refs. \cite{bucker2011twin,bucker2013vibrational}, a source for stimulated emission of matter waves in twin beams is created by transferring a condensate into the collective first-excited state. The typical decay rate of the system is 3 ms, so it is very important to find optimal controls that can transfer the condensate into the excited state faster than this decay rate and still allow time for subsequent experiments \cite{bucker2011twin,van2016optimal}. This condensate driving control problem has been investigated using a number of different QOC algorithms in Refs. \cite{van2016optimal,jager2014optimal,hohenester2007optimal}. We give a brief account of the condensate driving control problem and, for more details on the experimental setup we refer to Refs. \cite{van2016optimal,bucker2013vibrational,bucker2011twin,lesanovskyPotential}. In a mean-field treatment, the dynamics of the condensate are well described by an effective one-dimensional Gross-Pitaevskii equation (GPE),
\begin{align}
    i\diff{\psi}{t}&=-\frac{1}{2m}\diff{^2\psi}{x^2}+V(x,u)\psi+\beta |	\psi|^2 \psi \nonumber \\
    					&=\bigl(\hat{H}+\beta |\psi|^2\bigr) \psi, \label{GPE}
\end{align}
where $\hbar=1$, $\beta$ is the effective nonlinear self-interaction, and $\hat{H}$ is the Hamiltonian. The atom chip experiment is tightly confined along two transverse directions and weakly confined along the axial direction. One of the transverse directions is so strongly confining that the state is frozen into the ground state. The dynamics along the axial direction is slow compared with the other transverse direction, which allows for a description with an effective one-dimensional GPE with a nonlinear dependence of $\beta$ on the atom number \cite{van2016optimal,gerbier2004quasi}. For 700 atoms, one finds $\beta=1830 \hbar  \; \text{Hz} \mu \text{m}$ \cite{van2016optimal}.

The potential is an anisotropic Ioffe-Pritchard trap dressed by a radio-frequency potential \cite{van2016optimal,bucker2013vibrational,lesanovskyPotential}. In Ref. \cite{van2016optimal}, this potential is approximated by a polynomial,
\begin{equation}
    V(x,u(t))=p_2\bigl(x-u(t)\bigr)^2+p_4\bigl(x-u(t)\bigr)^4+p_6\bigl(x-u(t)\bigr)^6, \label{potential}
\end{equation}
where the control $u(t)$ is the displacement of the trap. The coefficients are given by $p_2 = 2\pi \hbar \cdot 310 r_0^{-2}\;\text{Hz}$, $p_4=2\pi\hbar \cdot 13.6 r_0^{-4}\;\text{Hz}$, and $p_6=-2\pi\hbar \cdot 0.0634 r_0^{-6}\;\text{Hz}$, with $r_0=172 \text{nm}$ \cite{van2016optimal}. For an in-depth discussion of the experimental setup we refer the reader to Refs. \cite{van2016optimal,bucker2011twin}. The goal is to transfer the initial state $\psi_0$ into the target state $\psi_t$ after a duration of \textit{T}. The initial state is the ground state for $u(t=0)=0$ and the target state is the first-excited state for $u(t=T)=0$. 

In order to find a control that transfers the initial state into the target state, the problem is expressed as a minimization of the cost functional $J(u,\psi)$,
\begin{equation}
J(u(t),\psi)=\frac{1}{2}\Bigl(1-|\langle \psi_t|\psi(T)\rangle|^2\Bigr)+\frac{\gamma}{2}\int_0^T\dot{u}(t)^2\text{d}t. \label{costFun}
\end{equation}
In the first term, $F=|\langle \psi_t|\psi(T)\rangle|^2$ is the fidelity and $1-F$ is the infidelity, which quantifies the difference between the final state and the target state \cite{mennemann2015optimal,hohenester2007optimal,von2008computational,jager2014optimal}. 
The second term in Eq. (\ref{costFun}) is the regularization that penalizes strong fluctuations in the control, which accounts for the fact that very fast changes cannot be realized experimentally. We found that $\gamma = 1 \cdot 10^{-6}$ gives acceptably smooth solutions.

Here we investigate the control problem at $T=1.09 \text{ms}$, which was reported in Ref. \cite{van2016optimal} to be the value of the quantum speed limit (QSL). The QSL is the lowest-duration \textit{T} where solutions above $F\geq 0.99$ can be found. In recent work, it has been shown that constrained control problems such as condensate driving become NP-hard close to the QSL \cite{bukovm,day2018glassy,bukov2018broken}. Hence, comparing the performance of the quantum control algorithms close to the QSL provides a stringent test of their individual performance. A solution at the QSL is also a solution at all longer durations, since the target state is an eigenstate of the GPE. Attempting to solve the problem at shorter durations can give improvements to the estimate of the QSL but it also requires systematic global exploration, which is discussed for the condensate driving problem in Ref. \cite{globalPaper}.

\section{Quantum Optimal Control} \label{sec:QuantumOptimalControl}
\begin{figure*}[t]
\begin{minipage}[t]{.48\textwidth}
\includegraphics[trim = 0mm 0mm 0mm 0mm, clip, width=0.96\textwidth]{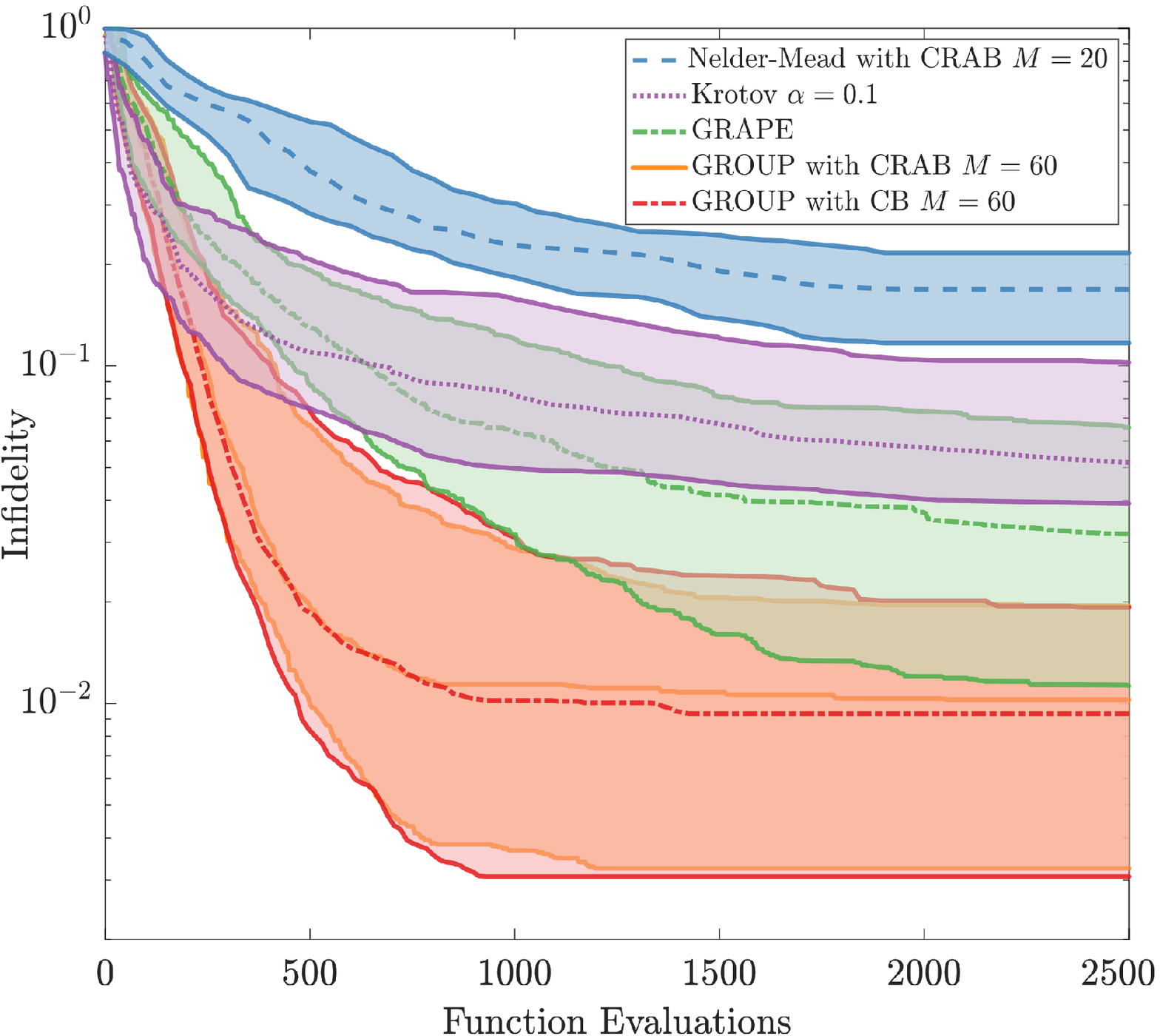}
\caption{(Color online) The infidelity as a function of the number of evaluations for each algorithm. One function evaluation is a solution of the GPE or Lagrange multiplier, given by Eq. (\ref{optim2}). The different algorithms are shown at the basis size or steps size where they performed the best (see legend). The dotted line shows the median and the shaded area indicates the 25\% and 75\% quartiles found from 100 different random initial controls. The quasi-Newton method \textsc{bfgs} was used together with \textsc{grape} and \textsc{group}.}
\label{fig:convergenceCompare}
\end{minipage}
\quad
\begin{minipage}[t]{.48\textwidth}
\includegraphics[width=\textwidth]{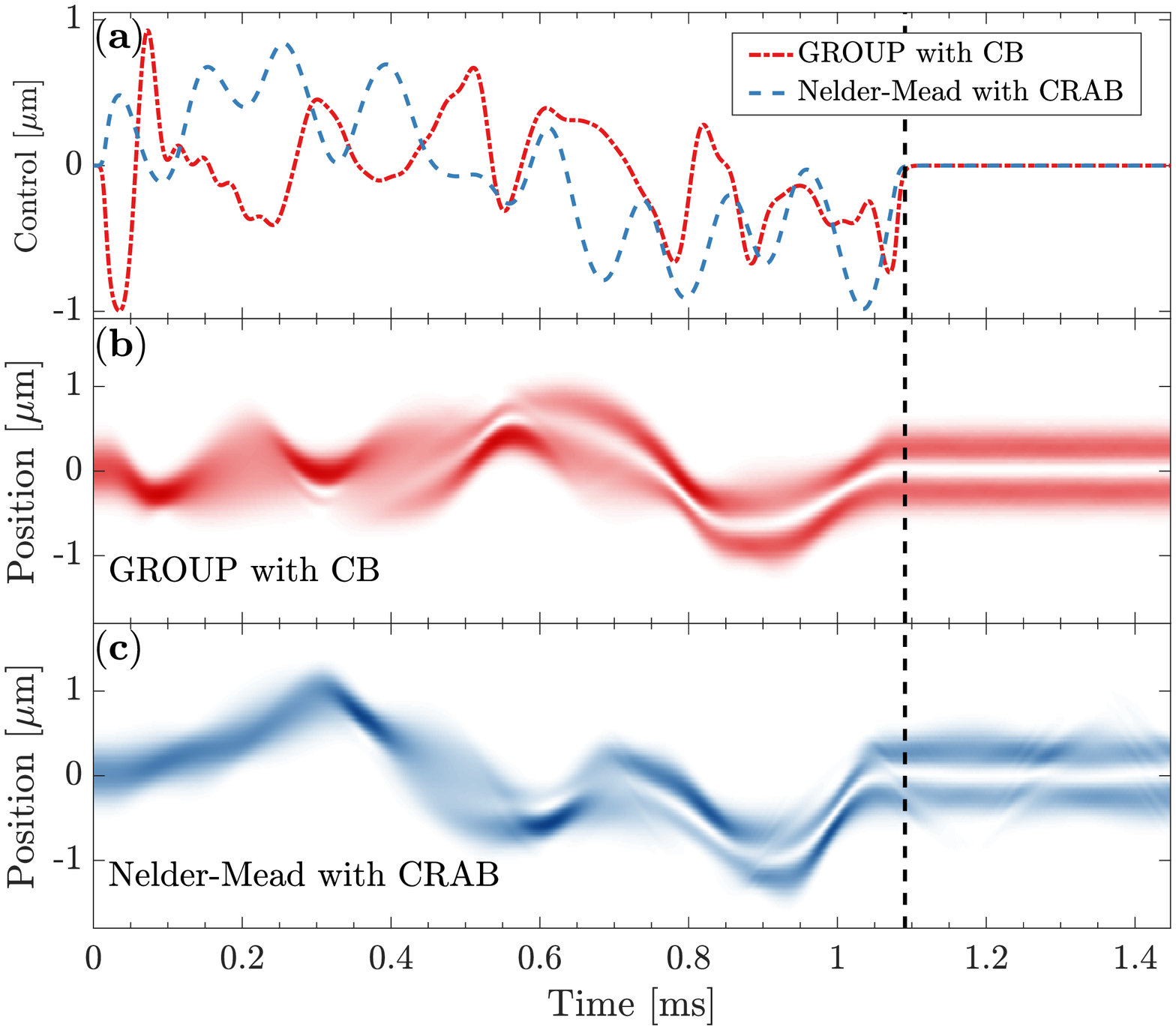}
\caption{(Color online) (\textbf{a}) The best-case controls found using \textsc{group} with \textsc{cb} (red) and Nelder-Mead with \textsc{crab} (blue) after 2500 iterations. The control is held constant after the vertical dashed line. (\textbf{b}) The density for the condensate ($|\psi(x,t)|^2$) found when using the solution from \textsc{group}. After the vertical line, the density is constant since the state has converged with $F=0.999$ to the first-excited state. (\textbf{c}) The same is shown as in (\textbf{b}) for Nelder-Mead with \textsc{crab}, which has residual oscillations since the final fidelity is $F=0.95$.}
\label{fig:controlCompare}
\end{minipage}
\end{figure*}
In QOC the goal is to minimize the cost functional $J(u,\psi)$ while satisfying the constraints from the GPE. The constraint can be handled using a Lagrange multiplier, 
\begin{align}
\mathcal{L}(\psi,u,\chi)=J+\Re \int_0^T \bigl\langle \chi \bigl|i\dot{\psi}-\hat{H}\psi - \beta |\psi|^2 \psi \bigr\rangle \text{d}t, \label{GRAPELagrangian}
\end{align}
where $\chi$ is a Lagrange multiplier \cite{peirce1988optimal}. Note, that the usual linear Schr\" odinger equation is a special case for $\beta=0$, so the methodology we present here can be directly extended to such systems. At a local minimum, all three variational derivatives $D_{\delta \chi} \mathcal{L}$, $D_{\delta \psi} \mathcal{L}$ and $D_{\delta u} \mathcal{L}$ are zero. This gives the following optimality equations (see Appendix \ref{sec:appEoO} for details):
\begin{align}
i  \dot{\psi} &= \hat{H} \psi+ \beta |\psi|^2 \psi, \label{optim1} \\
i  \dot{\chi} &= \bigl(\hat{H}+2\beta|\psi|^2\bigr)\chi+ \beta \psi^2\chi^*, \label{optim2} \\
\gamma \ddot{u}&=-\Re\biggl\langle \chi\biggl|\diff{\hat{H}}{u}\biggr|\psi\biggr\rangle, \label{optim3}
\end{align}
and the associated boundary conditions,
\begin{align}
\psi(0)&=\psi_0, \label{boundary1} \\
i \chi(T) &= - \langle \psi_t | \psi(T)\rangle \psi_t, \label{boundary2} \\
u(0)&=u_0, \quad u(T)=u_T. \label{boundary3}
\end{align}
Ideally, these equations would be solved analytically, which would directly give the optimal solutions. Unfortunately, such analytic solutions are, in general, infeasible so it is necessary to use iterative numerical algorithms \cite{hohenester2007optimal,mennemann2015optimal}.

When performing the numerical optimization of the cost functional, it is necessary to discretize the control in steps $\Delta t$, where $\Delta t$ is set by the required accuracy when numerically solving the GPE. This means that the control becomes represented by a vector of length $N=\floor{T/\Delta t}+1$. Typically, in \textsc{grape} and Krotov's method, the dimension for the space of admissible control ($M$) is the same as for the simulation $N=M$. However, with a proper change of basis, the optimal controls could adequately be described by another basis with much smaller dimension \cite{motzoi2011optimal,lloyd2014information}. In this sense, setting $N=M$ often introduces too many degrees of freedom for the control problem. In the simulations performed here, we have $N\simeq 3500$, whereas by a proper choice of basis, we have only $M\simeq 50$. These potential large reductions in the dimension of the control problem are the motivation for reduced or chopped basis methods. In recent years, \textsc{crab} has emerged as an attractive alternative to the conventional QOC methods since it parametrizes the control in a reduced-basis \cite{brouzos2015quantum,doria2011optimal,van2016optimal,caneva2011chopped}.

As mentioned, the main purpose of this paper is to introduce and numerically test \textsc{group}. We first give a brief review of \textsc{grape} and \textsc{crab} since \textsc{group} builds on these methods. When discussing these methods, one typically works with the reduced cost functional $\hat{J}(u)=J(u,\psi_u)$. Here, $\psi_u$ is the unique solution to the GPE, which is found by solving Eq. (\ref{GPE}) with $u(t)$ \cite{von2008computational}. Our objective is to find a minimum, and preferably the global minimum of $\hat{J}(u)$. 

\subsection{GRAPE}
A standard approach to minimizing $\hat{J}(u)$ is the \textsc{grape} algorithm \cite{khaneja2005optimal,jager2014optimal,mennemann2015optimal}. The simplest version of this method is to update the control along the gradient,
\begin{equation}
u^{(i+1)}=u^{(i)}-\alpha^{(i)} \nabla \hat{J}\bigl(u^{(i)}\bigr), \quad i=0,1,2,... \label{SD}
\end{equation}
Here the index \textit{i} refers to the current iteration. This is the steepest descent method. An appropriate value for  $\alpha$ is found using a step-size algorithm, which finds a satisfactory solution to the one-dimensional optimization problem,
\begin{equation}
\alpha^{(i)}=\argmin_\alpha \hat{J}\Bigl( u^{(i)}-\alpha \nabla \hat{J}\bigl(u^{(i)}\bigr) \Bigr).
\end{equation}
Note that in Eq. (\ref{SD}), the control is updated for all times $0\leq t \leq T$ at once. This makes \textsc{grape} a concurrent method, which is different from Krotov's method presented later where the control is updated sequentially for each time slice.

An important but subtle point is the use of the gradient $\nabla \hat{J}\bigl(u^{(i)}\bigr)$ in Eq. (\ref{SD}). The complication arises from the fact the the gradient is defined in a function space \textit{X}. In this space, the gradient is the unique element such that $(\nabla \hat{J},\delta u)_X=D_{\delta u} \hat{J}$ for all possible variations $\delta u$. Hence, the gradient depends on the choice of the inner product for the function space \textit{X}. This has already been discussed in a number of Refs. \cite{von2008computational,mennemann2015optimal}. Specifically, all variations $\delta u$ should satisfy the boundary condition in Eq. (\ref{boundary3}) giving $\delta u(0)=\delta u(T)=0$. A choice that will fulfill this requirement is the $H^1$-space with the inner product $(u,v)_{H^1}=\int_0^T \dot{u} \dot{v} \text{d}t.$ As shown in appendix \ref{sec:appGRAPE}, this choice gives the result,
\begin{equation}
\dif{^2}{t^2} [\nabla \hat{J}(u)] = \Re\biggl\langle \chi\biggl|\diff{\hat{H}}{u}\biggr|\psi\biggr\rangle + \gamma \ddot{u} \label{H1Grad}
\end{equation}
where $\psi$ and $\chi$ are the solutions of Eq. (\ref{optim1}) and Eq. (\ref{optim2}). This is a Poisson equation for the control in time so we can choose the Dirichlet boundary conditions $[\nabla \hat{J}(u)](0)=\nabla [\hat{J}(u)](T)=0$.  These conditions imply that \textsc{grape} preserves the boundary condition in Eq. (\ref{boundary3}) in any iteration. As pointed out in in a number of Refs. \cite{von2008computational,mennemann2015optimal} if we had made the canonical choice of $X=L^2$, then the gradient would not vanish at the boundary. In this case, the boundary condition must be enforced numerically by simply setting the gradient to be zero at the boundary, which can greatly decrease the performance of the algorithm.

The update written in Eq. (\ref{SD}) is the steepest descent algorithm, which is also illustrated in Fig. \ref{fig:cartoon}. As shown in this figure, the optimization can be improved by using a quasi-Newton method such as \textsc{bfgs} instead of steepest descent. Again, special care has to be taken when working in $H^1$-space, and we use the matrix free version of \textsc{l-bfgs} described in Ref. \cite{von2008computational}.

\subsection{Chopped Basis and CRAB}
\begin{figure*}[t]
\begin{minipage}[t]{.48\textwidth}
  \includegraphics[width=\textwidth]{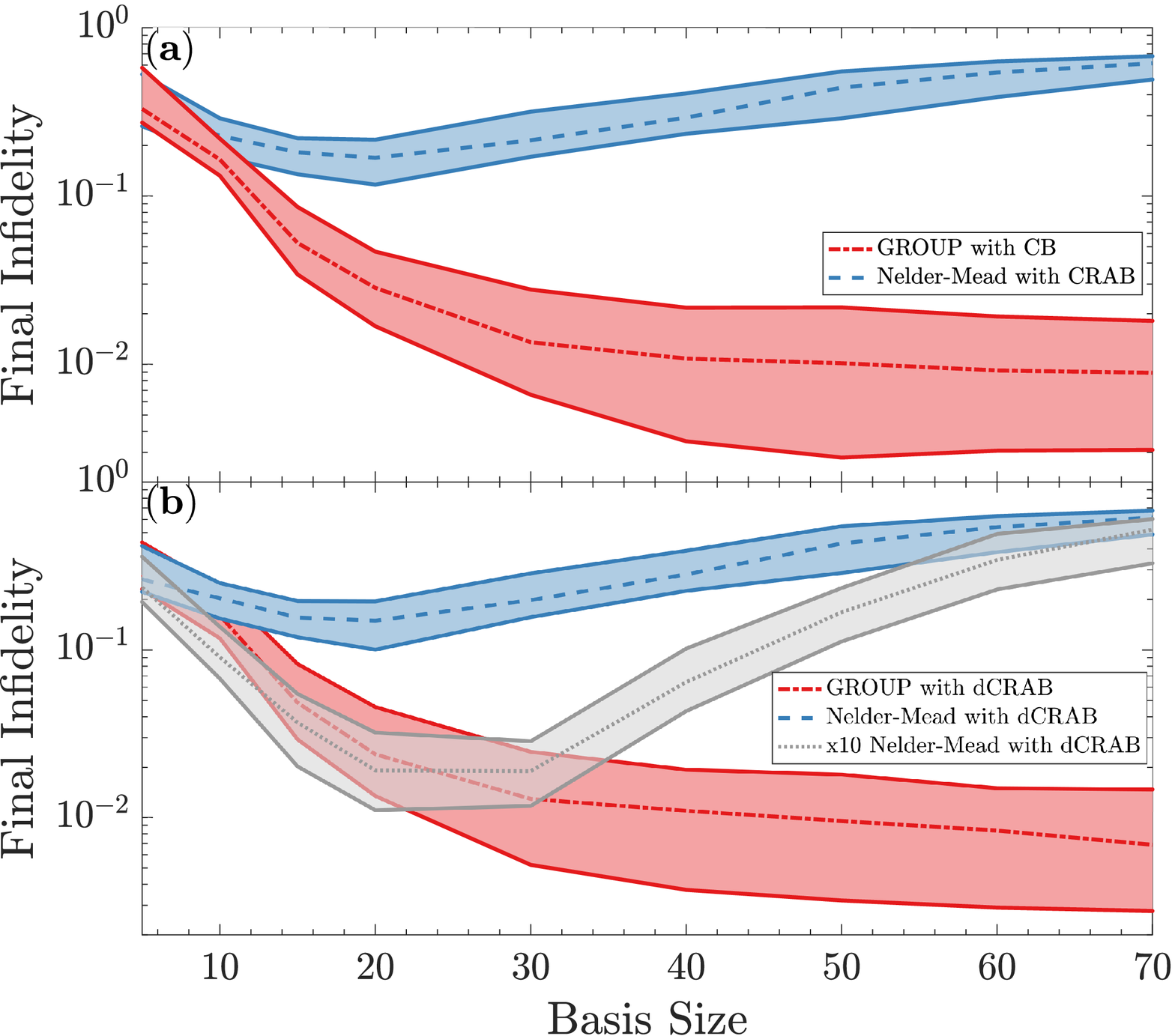}
  \caption{(Color online) (\textbf{a}) The final infidelity after 2500 function evaluations for \textsc{group} with \textsc{cb} (red) and Nelder-Mead with \textsc{crab} (blue) for different basis sizes. The dotted line is the median and the shaded area display the 25\% and 75\% quartiles found from the 100 different random initial controls. (\textbf{b}) The same as (\textbf{a}) for Nelder-Mead and \textsc{group} with dressed \textsc{crab}. The gray plot shows the final infidelity for Nelder-Mead with dressed \textsc{crab} after 25,000 iterations rather than 2500 iterations.}
\label{fig:basisCompare}
\end{minipage}
\quad
\begin{minipage}[t]{.48\textwidth}
  \includegraphics[width=\textwidth]{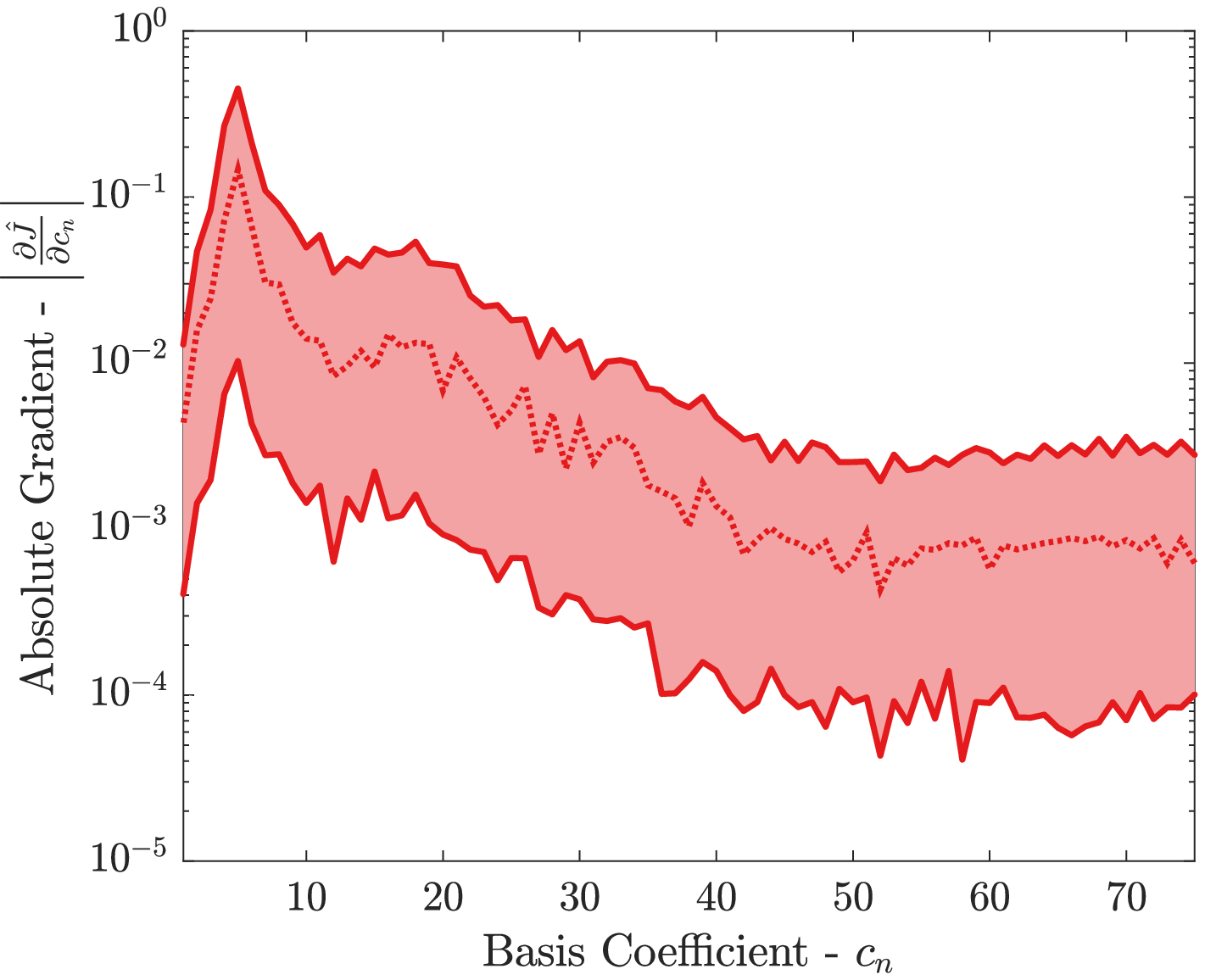}
  \caption{The absolute value of the partial derivative with respect to basis coefficients $|\partial \hat{J}(\mathbf{c})/\partial c_n|$ found using Eq. (\ref{groupGRAD}). The dotted line shows the median of the 100 gradients taken at the start of the optimization. The shaded area indicates the 25\% and 75\% quartiles.}
  \label{fig:gradient}
\end{minipage}
\end{figure*}
As discussed above, the control is represented by a vector of length $N=\floor{T/\Delta t}+1$ due to the temporal discretization. In \textsc{grape} the dimensionality of the space of admissible controls ($M$) is $N=M$. However, by expanding the control in a proper basis it is possible to substantially reduce the dimension of the control problem,
\begin{equation}
u(t) = u_0(t) + S(t) \sum_{n=1}^M c_n f_n(t), \label{crabExpansion}
\end{equation}
where the $f_n$'s are the basis functions. $S(t)$ is a shape function that enforces the boundary condition in Eq. (\ref{boundary3}), which gives $S(0)=S(T)=0$. In this chopped basis, we optimize the coefficients $c_n$ instead of full control $u(t)$, which implies that the cost function is $\hat{J}(\mathbf{c})$ where $\mathbf{c}=(c_1,c_2,...,c_M)$. This method is known as a chopped basis - \textsc{cb}. The functions $f_n$ must be chosen sensibly based on physical insight, which would typically be sinusoidal functions around characteristic frequencies. 
The reduced dimension of $\hat{J}(\mathbf{c})$ enables the use of the gradient-free Nelder-Mead algorithm. Gradient-free methods have the advantage that there is no need to implement code that calculates the gradient by solving, e.g., Eq. (\ref{H1Grad}), which also requires solving the equation for the Lagrange multiplier given by Eq. (\ref{optim2}). This is particularly an advantage whenever calculation of the gradient is infeasible or too resource consuming \cite{doria2011optimal}.

For the control problem discussed here, we found the following expansion useful:
\begin{equation}
u(t)=u_0(t)+S(t) \sum_{n=1}^M c_n \sin\biggl(\frac{\omega_n t}{T}\biggr), \label{ourBasis}
\end{equation}
where $\omega_n = n\pi$ is a set of frequencies. This type of chopped basis was extended in Ref. \cite{caneva2011chopped} by introducing the chopped random basis or \textsc{crab}. In \textsc{crab} the frequencies $\omega_n$ are randomly shifted as $\omega_n= (n+r_n)\pi$, where $-0.5\leq r_n \leq 0.5$ are initially chosen random numbers. The optimization is repeated a number of times with different values of $r_n$'s. This is a central idea in \textsc{crab}, since it allows the algorithm to explore different basis functions with slightly similar frequencies starting from the same $u_0$. An optimization within a \textsc{crab} can principally be done using any method, but it is typically done using the gradient-free method Nelder-Mead \cite{doria2011optimal,caneva2011chopped,van2016optimal}.

\subsection{GROUP}
In \textsc{group} the best features of the two previous methods are combined. We parametrize the control in some basis as in Eq. (\ref{crabExpansion}). However, instead of using a gradient-free method to search in the chopped basis, we use the gradient-descent methods from \textsc{grape}. Here the gradient is with respect to the expansion coefficients [$\nabla \hat{J}(\mathbf{c})$]. Given this gradient an iterative update analogous to Eq. (\ref{SD}) can be directly applied. Just as in \textsc{grape} we need to find an analytic expression for the gradient similar to Eq. (\ref{H1Grad}). The partial derivative of $\hat{J}(\mathbf{c})$ with respect to $c_n$ can be found using the chain rule for variational derivatives (see Appendix \ref{sec:appGROUPgrad} for more details). The result is
\begin{align}
\diff{\hat{J}(\mathbf{c})}{c_n}&=D_{S(t)f_n(t)}\hat{J}(u) \nonumber \\
&=-\int_0^T \biggl(\Re \biggl\langle \chi\biggl|\diff{\hat{H}}{u}\biggr|\psi\biggr\rangle + \gamma \ddot{u} \biggr)S(t)f_n(t)\text{d}t. \label{groupGRAD}
\end{align}
Note that these partial derivatives only differ in the $f_n(t)$ function in the integrand. This expression is valid for any \textsc{cb} and \textsc{crab}. The quantity in the bracket only needs to be computed once, which contains $\psi$ and $\chi$ that are found by the numerically expensive solution of Eqs. (\ref{optim1}) and (\ref{optim2}). The cost of calculating all the partial derivatives and hence the full gradient [$\nabla \hat{J}(\mathbf{c})$] is dominated by the solution of  Eqs. (\ref{optim1}) and (\ref{optim2}), which implies that the time needed for calculating the gradient in \textsc{group} is comparable to \textsc{grape}.

In the comparative numerical studies presented below we performed the \textsc{group} optimization using the quasi-Newton method \textsc{bfgs}. Note that when optimizing $\hat{J}(\mathbf{c})$, the optimization is done in the usual $l^2$- and not $H^1$-space, so all the standard methods for \textsc{bfgs} can be directly applied.

All numerical \textsc{group} results in this paper were obtained using the formalism from Eq. (\ref{groupGRAD}) extended with a filter function. In Ref. \cite{machnes2015gradient}, it was proposed to compute the gradient in a reduced-basis $\partial \hat J(\mathbf{c})/\partial c_n$ but using another method named gradient optimization of analytic controls (\textsc{goat}). We give a brief account of how this method can be extended to the condensate driving control problem in in Appendix \ref{sec:appGOAT}. Briefly, the \textsc{goat} algorithm is more numerically expensive than \textsc{group} but could potentially offer advantages if ultra low infidelities are required.

\subsection{Krotov's Method}
Krotov's method is an alternative to the standard Lagrange multiplier method used in \textsc{grape} and \textsc{group} \cite{konnov1999global}. In Krotov's method, the cost functional given by Eq. (\ref{costFun}) is rewritten so the GPE appears explicitly and conditions for a guaranteed decrease in the cost are directly built in \cite{konnov1999global,sklarz2002loading,reich2012monotonically,jager2014optimal}. This allows Krotov's method to give an optimal control algorithm that directly ensures a monotonic decrease in the cost and it is expected to provide fast convergence. The resulting update for the control is (see Appendix \ref{sec:appKrotov} for more details)
\begin{equation}
u^{(i+1)}(t)=u^{(i)}(t)+\alpha S(t) \Re \biggl\langle \chi^{(i)}(t)\biggl|\diff{\hat{H}}{u}\biggr|_{u^{(i)}}\biggr|\psi^{(i+1)}(t)\biggr\rangle. \label{krotovUpdate}
\end{equation}
Here, $\psi^{(i)}$ and $\chi^{(i)}$ are the solutions to the GPE Eq. (\ref{optim2}) and the Lagrange multiplier Eq. (\ref{optim1}) from \textsc{grape}. $S(t)$ is again a shape function $0\leq S(t) \leq 1$ that turns the control off at $t=0$ and $t=T$, which ensures that the boundary condition for the control is always satisfied in Eq. (\ref{boundary3}). $\alpha$ is the step-size parameter that must be selected for proper convergence.  Note that both the current iteration index \textit{i} and the next index $i+1$ appear in the equation. Specifically, the control at the next iteration $u^{(i+1)}$ depends on the states in the next iteration $\psi^{(i+1)}$. This makes Krotov's method a sequential algorithm where the next control $u^{(i+1)}$ is being calculated while the equations of motion are being solved along that control. This is very different from the other methods presented here, where the control is updated concurrently for all values of $0< t < T$. Note that since the GPE is nonlinear, in the states a monotonic decrease in the cost can only be guaranteed if Eq. (\ref{krotovUpdate}) includes an additional term that is proportional to the difference in the states between iterations \cite{sklarz2002loading,reich2012monotonically,jager2014optimal}. However, for the small values of $\beta$ discussed here neglecting this term does not notably affect the monotonic decrease in the cost \cite{jager2014optimal}. The derivative is typically with respect to the next iteration $(u^{(i+1)})$, but for the small values of $\alpha$ used here it is acceptable to use the current iteration $(u^{(i)})$ \cite{jager2014optimal,reich2012monotonically}.

\section{Filter Function} \label{sec:filter}
\begin{figure*}[t]
\begin{minipage}[t]{.48\textwidth}
  \includegraphics[width=\textwidth]{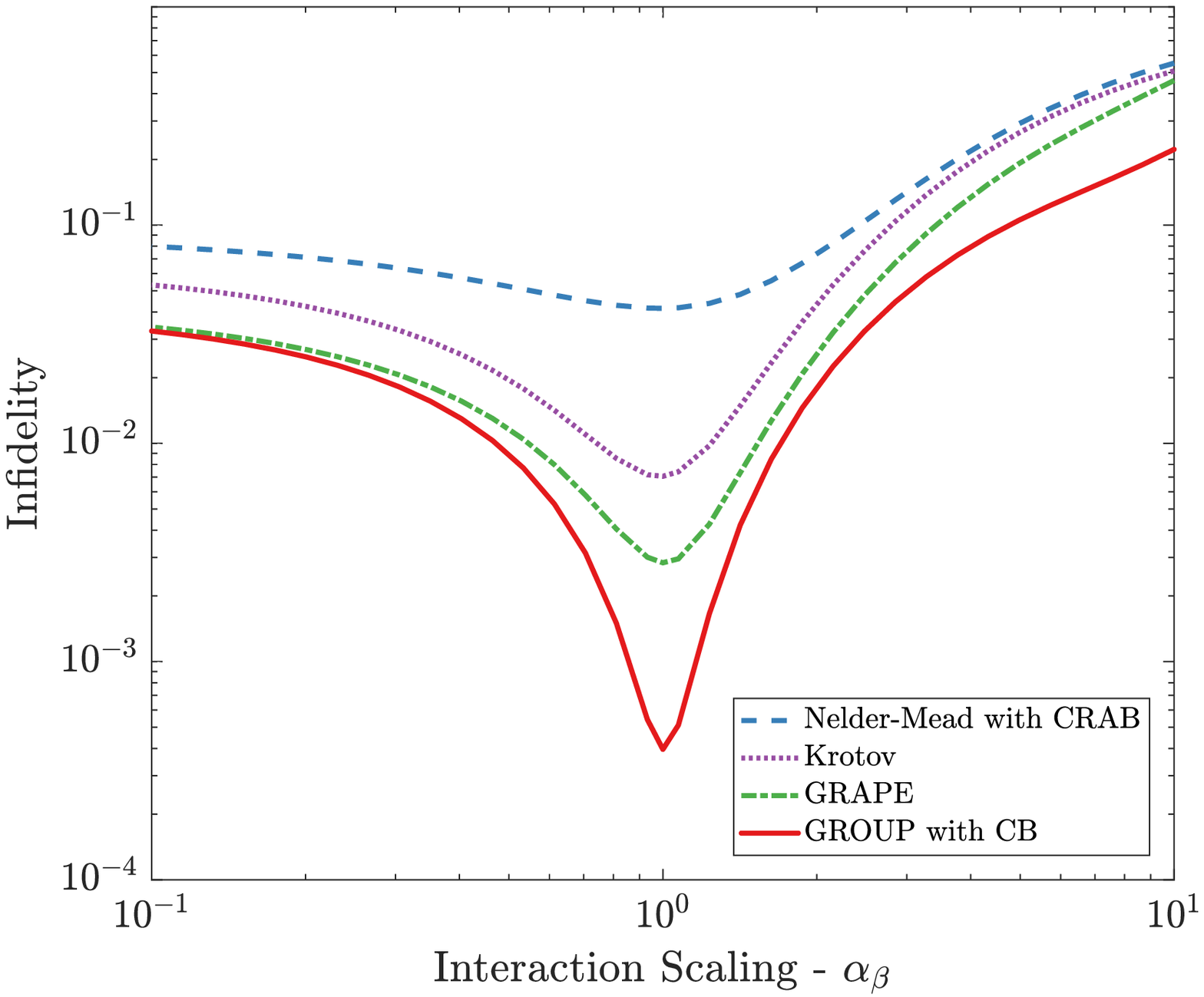}
  \caption{The robustness of the optimal solutions from each algorithm to rescaling of the self-interaction in Eq. (\ref{GPE}): $\tilde{\beta}=\alpha_\beta \beta$, where $\alpha_\beta$ is the scaling.}
  \label{fig:betaRobustness}
\end{minipage}
\quad
\begin{minipage}[t]{.48\textwidth}
  \includegraphics[width=\textwidth]{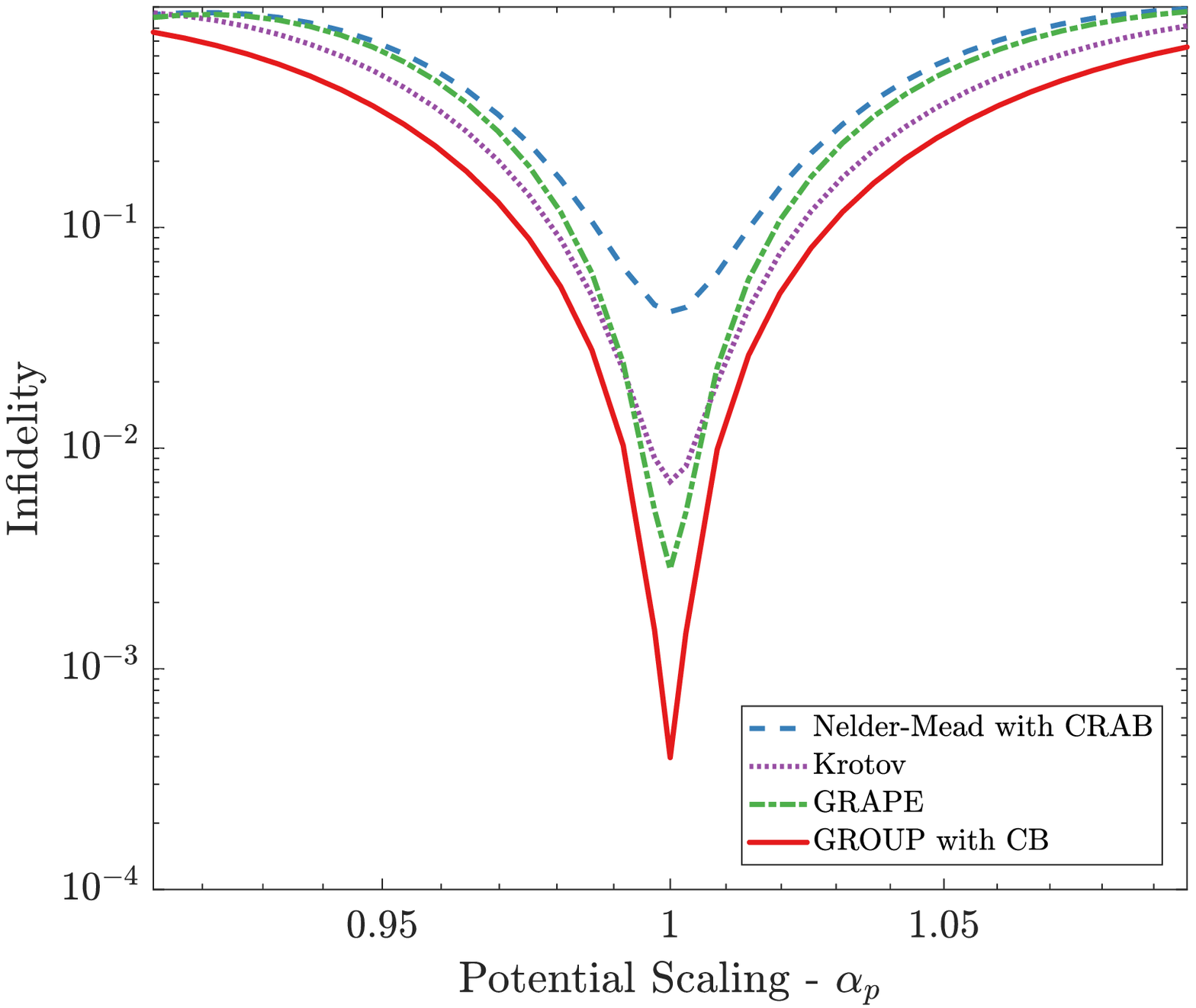}
  \caption{The robustness of the optimal solutions from each algorithm to rescaling of the potential coefficients in Eq. (\ref{potential}): $\tilde{p}_i=\alpha_p p_i,\;i=2,4,6$, where $\alpha_p$ is the scaling.}
\label{fig:pScaleRobustness}
\end{minipage}
\end{figure*}
In order to obtain a close match to the experimental conditions in the condensate driving control problem, it is also necessary to include the finite bandwidth of the control electronics. This effect causes the control $u(t)$ to become distorted into $v(t)$ and the atoms experience the potential from $V(x,v)$ \cite{van2016optimal}. The distortion is large enough to cause the fidelity to drop by a couple of percent, so it must be included in the modeling. It has previously been discussed how to include this type of effect into \textsc{grape} \cite{jager2013optimal} and Krotov-type methods \cite{reich2014optimal,gollub2008monotonic}. The distorted control is given by a convolution with the filter $h(\tau)$
\begin{equation}
v(t) = (h*u)(t) = \int_0^t h(\tau) u(t-\tau)\text{d}\tau.
\end{equation}
  
The presence of the filter changes the expressions for the gradients (see Appendices \ref{sec:appGRAPE} and \ref{sec:appGROUPgrad} for more details). The \textsc{grape} gradient [Eq. (\ref{H1Grad})] becomes
\begin{align}
\dif{^2}{t^2} [\nabla \hat{J}(u)] &= -\gamma \dot{v}(T) h(T-t) \nonumber \\
&+\int_t^T \biggl(\Re \biggl\langle \chi\biggl| \diff{\hat{H}}{v}\biggr|\psi\biggr\rangle + \gamma \ddot{v}\biggr)h(t'-t)\text{d}t'
\end{align}
Similarly, the expression for the \textsc{group} gradient becomes 
\begin{align}
\diff{\hat{J}(\mathbf{c})}{c_n}&=-\int_0^T \biggl(\Re\biggl\langle \chi\biggl|\diff{\hat{H}}{v}\biggr|\psi\biggr\rangle + \gamma \ddot{v} \biggr)(h*Sf_n)(t)\text{d}t \nonumber \\
&+\gamma \dot{v}(T) (h*Sf_n)(T) \label{groupGRADfilter}
\end{align}
It is not straightforward to include such a filter function in Krotov's method \cite{reich2014optimal}. For the comparison below, the simulations for Krotov are performed without the filter \cite{reich2014optimal}. The filter gives rise to another potential complication, which is that although $u(T)=0$, it can occur that $v(T)\neq0$. This could slightly perturb the final state, since the control cannot be instantaneously quenched to zero due to the filter. For the optimal solutions found here, this effect did not notably affect the fidelity.

\section{Numerical Results} \label{sec:Results}
In the last section, we gave an introduction to four different quantum control algorithms. In this section, we will compare them numerically when applied on condensate the driving control problem.

\subsection{Convergence Behavior}
We applied the following QOC algorithms \textsc{grape}, Krotov, Nelder-Mead with \textsc{crab} and \textsc{group} with \textsc{cb} and \textsc{crab} to the condensate driving control problem. We applied the algorithms to the same 100 initial controls, which were randomly generated using Eq. (\ref{crabExpansion}). The convergence behavior of the different methods is illustrated in Fig. \ref{fig:convergenceCompare}. Here the median and 25\% and 75\% quartiles are shown for the different algorithms, which gives an impression of the expected behavior for each method on this problem. One function evaluation is a solution of the GPE or the equation for the Lagrange multiplier [Eq. (\ref{optim2})]. 

Throughout the optimization \textsc{group} achieves the lowest infidelities. At the end of the optimization \textsc{group} has the best infidelity followed by \textsc{grape} and Krotov. There is no particular difference between \textsc{group} using \textsc{cb} or \textsc{crab} so we will refer to them both as \textsc{group}. Nelder-Mead using \textsc{crab} has the slowest convergence rate of the four methods, since it does not utilize derivative information. This is in accordance with the picture presented in Fig. \ref{fig:cartoon}, which shows that derivative-based methods are typically faster than derivative-free. 

The optimization curves in Fig. \ref{fig:convergenceCompare} can be split into two regimes: one at a high number of function evaluations after 600 and one below. Below 600 evaluations, the three derivative-based methods have similar rates of convergence, but \textsc{group} and Krotov perform better than \textsc{grape}. In the high number of function evaluations regime \textsc{group} performs better than \textsc{grape} and Krotov. We attribute this to the fact that the basis gradually steers \textsc{group} towards a more profitable part of the optimization landscape. This shows the complexity of the optimization landscape, since although the algorithms start at the same point and perform local greedy optimization they converge towards different points with different infidelities. This is different from the situation in Fig. \ref{fig:cartoon} where all the algorithms converge to the same point since the two-dimensional landscape is much simpler than the control problem, which has dimension $M\simeq 3500$ or $M\simeq 50$.

In the high function evaluations regime Krotov and \textsc{grape} switch places and \textsc{grape} finds better infidelities than Krotov. These results show that Krotov achieves fast initial reductions in the infidelity, but it slows considerably down when approaching the optimum. \textsc{grape} does not exhibit this behavior, which we attribute to the Hessian approximation from \textsc{bfgs}, since the cost function can be well described by a second-order expansion close to the optimum \cite{nocedal2006numerical}. Similar results were also reported in Ref. \cite{jager2014optimal}. In principle, Krotov's method can also be combined with a \textsc{bfgs} type method \cite{eitan2011optimal}, but it has been reported in Refs. \cite{eitan2011optimal,jager2014optimal} that this does not significantly improve the convergence.

This comparison focuses on the expected behavior. Each of the methods have a few optimization runs that perform significantly better, which reflects that each algorithm has some specific seeds where it just happens to search the optimization landscape in the most favorable manner. \textsc{group} had the individual optimization runs with the lowest infidelity.

The best controls found for \textsc{group} with \textsc{cb} and Nelder-Mead with \textsc{crab} are shown in Fig. \ref{fig:controlCompare}, with respectively, $F=0.999$ and $F=0.95$. This figure also shows the density for the condensate when propagated along the optimal controls. After \textit{T}, the control is held constant at $u=0$. The Nelder-Mead with \textsc{crab} solution has a residual oscillation after \textit{T} due to residual excited-states' components in the solution.

\subsection{Reduced-Basis Size and Dressed Methods}
\begin{figure*}[t]
\begin{minipage}[t]{.48\textwidth}
  \includegraphics[width=\textwidth]{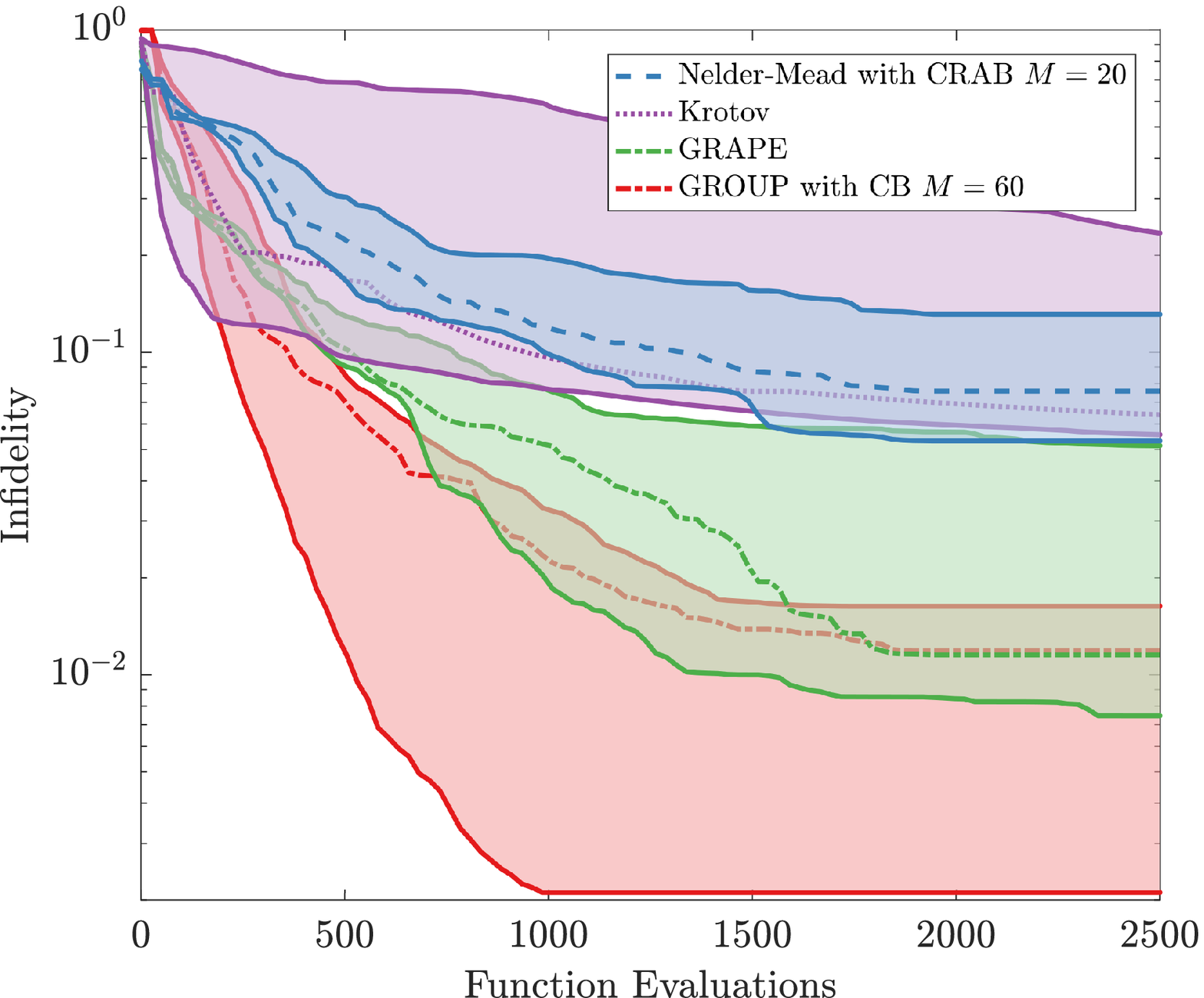}
  \caption{(Color online) Comparison of the convergence behavior for different values of the self-interaction for each algorithm. The optimizations are performed for 35 different scaled values of the self-interaction $\tilde{\beta}=\alpha_\beta \beta$, where $\alpha_\beta$ is the scaling. The dotted line shows the median and the shaded area indicates the 25\% and 75\% quartiles. The quasi-Newton method \textsc{bfgs} was used together with \textsc{grape} and \textsc{group}.}
  \label{fig:betaScaleCompare}
\end{minipage}
\quad
\begin{minipage}[t]{.48\textwidth}
  \includegraphics[width=\textwidth]{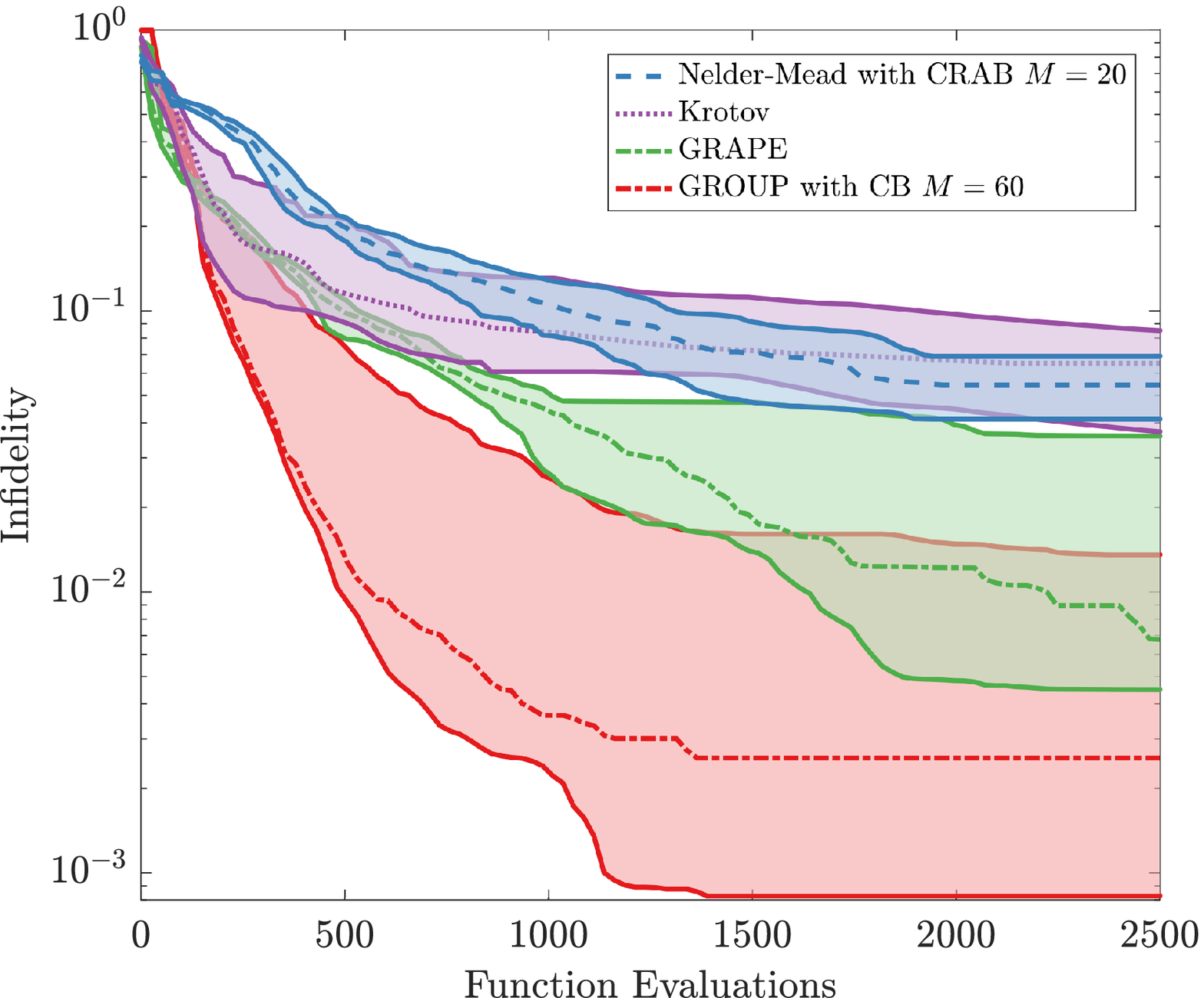}
  \caption{(Color online) Comparison of convergence behavior for different values of the potential coefficients. The optimizations are performed for 35 different scaled values of the potential coefficients $\tilde{p}=\alpha_p p_i$ with $i=2,4,6$, where $\alpha_p$ is the scaling. The dotted line shows the median and the shaded area indicates the 25\% and 75\% quartiles. The quasi-Newton method \textsc{bfgs} was used together with \textsc{grape} and \textsc{group}.}
\label{fig:potentialScaleCompare}
\end{minipage}
\end{figure*}
A limitation of the reduced-basis methods is that the optimization might get caught in an artificial trap introduced by limited bandwidth of the parametrization. As an example, consider the expansion in Eq. (\ref{ourBasis}) at some fixed \textit{M}. If the optimal solution requires frequencies above \textit{M}, then the optimization can never converge to this solution and it will get caught in an artificial trap. This effect would favor larger values of \textit{M}. On the other hand, increasing \textit{M} also increases the dimension of the control problem, which is exactly what the reduced-basis methods attempt to avoid. 

In Fig. \ref{fig:basisCompare}(\textbf{a}) we compare the trade-off between the risk of artificial traps and retaining a low dimension for Nelder-Mead with \textsc{crab} and \textsc{group} with \textsc{cb}. As expected, a too small basis $M\leq 10$ gives poor results for both algorithms since the optimal controls cannot be adequately described with these small basis sizes. Nelder-Mead with \textsc{crab} clearly has an optimal basis size around $M=20$ and it becomes worse with a larger dimension. This is due to the fact that the Nelder-Mead algorithm cannot effectively search within a large dimension. Surprisingly, the performance in \textsc{group} is very robust with respect to the basis size, and it even seems to prefer large basis sizes. We attribute this result to the fact that the large frequencies have small weights in the gradient, meaning that they contribute little to the search. This type of behavior is illustrated in Fig. \ref{fig:gradient}, which shows the median initial gradient for the 100 initial points. Above $n=50$, the partial derivatives are much smaller than for $n=5$, and hence it is the low frequencies that dominate the search.

A solution to the artificial trap problem was proposed in Ref. \cite{rach2015dressing} where a so-called dressed \textsc{crab} (d\textsc{crab}) was introduced. In d\textsc{crab} the optimization is restarted with new basis functions from the last optimum ($u^{(j-1)}$),
\begin{equation}
u^{(j)}=u^{(j-1)}+S(t)\sum_{n=1}^M c_n^{(j)}f_n^{(j)}\bigl(t,r_n^{(j)}\bigr).
\end{equation}
These iterations are known as superiterations \cite{rach2015dressing}. The new basis functions $f_n^{(j)}\bigl(t,r_n^{(j)}\bigr)$ are found by reselecting the random frequency shifts ($r_n$) in \textsc{crab} giving new basis functions in each superiteration. The initial coefficients are $c_n^{(j)}=0$ for all \textit{n} so the optimization starts from the previous control. The new basis functions give the algorithm the possibility to escape the artificial traps \cite{rach2015dressing}. Clearly, this formalism cannot be used in a normal chopped basis (\textsc{cb}). We apply the same methodology to \textsc{group} by simply following the same procedure for the superiterations and calculate the gradients using Eq. (\ref{groupGRADfilter}).

In Fig. \ref{fig:basisCompare}(\textbf{b}), we compare the effect of dressing for different basis sizes. \textsc{group} in combination with d\textsc{crab} (d\textsc{group}) only does slightly better than \textsc{group} with \textsc{cb}, which indicates that the solutions were already highly optimal after 2500 evaluations. We observe the same behavior for Nelder-Mead with d\textsc{crab} after 2500 evaluations. However, if Nelder-Mead with d\textsc{crab} is allowed to perform 25,000 evaluations, then it finds much better results especially for basis sizes around $M=20$. This shows that for low basis sizes, Nelder-Mead with d\textsc{crab} can escape the artificial traps and greatly improve the infidelity with enough function evaluations. The Nelder-Mead with d\textsc{crab} using 25,000 evaluations finds similar infidelities to \textsc{group} with d\textsc{crab} using 2500 evaluations, which shows that if Nelder-Mead runs for longer times, it can find results similar to gradient descent. At large basis sizes, d\textsc{crab} does not notably improve the infidelity, which we interpret as the optimization being blocked by Nelder-Mead's inability to efficiently search in high-dimensional landscapes rather than artificial traps.

\subsection{Robustness Analysis}
The solutions presented in this paper can only be directly applied in an experiment if they are stable against unavoidable experimental fluctuations in the system parameters.  As reported in Ref. \cite{van2016optimal}, the primary experimental fluctuations occur in the self-interaction ($\beta$) in Eq. (\ref{GPE}) due to variations in the atom number. This gives variations in $\beta$ below 14\%. For this reason, we have investigated the stability of the solutions against scaling of the self-interaction. In Fig. \ref{fig:betaRobustness}, the infidelity is calculated with the best solution from each algorithm using $\tilde{\beta}=\alpha_\beta \beta$, where $\alpha_\beta$ is the scaling factor. The solutions are generally stable with respect to fluctuations in the self-interaction, since atom interactions have little influence over the relative short duration investigated here. The \textsc{group} solution clearly exhibits superior behavior over the entire range of parameter variation. The \textsc{group} solution remains above $F=0.99$ in the range $0.5 \leq \alpha_\beta \leq 1.5$, which is above the experimental fluctuations reported in Ref. \cite{van2016optimal}. 

We have also investigated the robustness of the solutions against perturbations in the potential. In Fig. \ref{fig:pScaleRobustness}, the infidelity is calculated for each algorithm's best solution with rescaled potential coefficients ($\tilde{p}_i=\alpha_p p_i$) in Eq. (\ref{potential}). The solutions are much more sensitive to perturbations in the potential compared with the self-interaction.

We note, that it would most likely be possible to find much more stable solutions by including the stability directly in the optimization. This could, for example, be achieved by optimizing over an ensemble of different scaled interactions and potentials and defining the average cost over the ensemble.

\subsection{Convergence Behavior at Different Parameter Values}
The results presented in this section have so far only discussed a single choice of the system parameter values, meaning that the underlying optimization problem has been the same. The performance of an optimization algorithm may depend on the choice of problem, so in this section we investigate the performance of each algorithm for different problems by rescaling the system parameters. In Fig. \ref{fig:convergenceCompare}, we compared the convergence behavior for a number of different seeds but fixed system parameters. In Fig. \ref{fig:betaScaleCompare} Fig. \ref{fig:potentialScaleCompare}, we hold the seed fixed and vary the system parameters examined in the previous section. The seeds are those that gave the best result for each algorithm in the previous analysis. As in Fig. \ref{fig:convergenceCompare}, we have performed the analysis for different basis sizes and step sizes and we present the best case behavior in Figs. \ref{fig:betaScaleCompare} and\ref{fig:potentialScaleCompare}.

In Fig. \ref{fig:betaScaleCompare}, we compare the convergence behavior of each algorithm when optimizing with 35 different values of $\tilde{\beta}=\alpha_\beta \beta$ ranging from $\alpha_\beta = 0.1$ to $\alpha_\beta = 10$ as in the previous section. Although \textsc{grape} and \textsc{group} have similar medians for their final infidelities, \textsc{group} has a better overall convergence behavior. Both \textsc{group} and \textsc{grape} have faster rates of convergence than Nelder-Mead with \textsc{crab} and Krotov's method. In Fig. \ref{fig:convergenceCompare}, Krotov's method has a better final infidelity than Nelder-Mead with \textsc{crab}. However in Fig. \ref{fig:betaScaleCompare}, Nelder-Mead with \textsc{crab} catches up with Krotov after a high number of of function evaluations. In general, Krotov's method is struggling for the high values of $\beta$, which we attribute to the lack of including the term proportional to the difference in Eq. (\ref{krotovUpdate}). We believe that correctly including this term could substantially improve the convergence rate of Krotov's method. Note that here we present the results with the best basis for Nelder-Mead after optimizing over a range of basis sizes. The performance is significantly worse for nearby basis sizes 15 and 30.

In Fig. \ref{fig:potentialScaleCompare}, we perform a similar analysis for each algorithm just optimizing 35 different values of the potential with rescaled coefficients $\tilde{p}_i=\alpha_p p_i$, with $i=2,4,6$ ranging from $\alpha_p = 0.91$ to $\alpha_p = 1.09$ as in the previous section. Again, we observe that \textsc{group} followed by \textsc{grape} finds the solutions with the lowest infidelities. Krotov's method has fast initial reductions in the infidelity below 600 function evaluations, but it slows down above 1000 function evaluations. In this case, we also observe that Nelder-Mead with \textsc{crab} catches up with Krotov's method after a high number of function evaluations. Similarly to before, we observe in our other optimizations that Nelder-Mead has significantly worse performance for the nearby basis sizes 15 and 30.

We have also performed a similar analysis for rescaled values of the regularization factor ($\gamma$) in Eq. (\ref{costFun}), which showed that the regularization factor only has a minor impact on the results for all algorithms.

\section{Conclusion \& Outlook} \label{sec:Conclusion}
We have presented in detail how to perform gradient-based optimal control in a reduced-basis in \textsc{group}, which combines the advantages from \textsc{crab} and \textsc{grape} methods. We have presented a benchmark with other quantum optimal control methods close to the quantum speed limit in the condensate driving control problem. Here, \textsc{group} type methods are competitive with other quantum control algorithms. We have also presented how to extend \textsc{group} type methods with experimentally motivated filter functions. It would be very interesting to compare these methods on other control problems in order to better understand the advantages and disadvantages of each method.
The formalism for \textsc{group} type methods presented here can also be applied to systems with a linear Schr\" odinger equation such as control of many-body systems \cite{doria2011optimal,van2016optimal}.  
We believe our method and analysis is a relevant addition to the repository of quantum control algorithms.

\section{Acknowledgements}
This work has been supported by the European Research Council and the Lundbeck Foundation. We would also like to thank Jesper H. M. Jensen for illuminating discussions and help with adjusting the manuscript. Antonio Negretti and Marie Bonneau helped with supplying the filter function and clarifying details in the modeling of the experiment.

\bibliographystyle{apsrev4-1}
\bibliography{references}

\appendix*
\section{Variations and Gradients}
\subsection{Equations of Optimality} \label{sec:appEoO}
Here we give a brief derivation of the optimality equations presented in the main text [Eqs. \ref{optim1})-(\ref{boundary3})]. The presentation here is based on Refs. \cite{mennemann2015optimal,hohenester2007optimal,jager2014optimal,von2008computational}. In order to find equations for an optimum that also satisfies the GPE, we need to calculate the variations with respect to the Lagrangian
\begin{equation}
\mathcal{L}(\psi,u,\chi)=J(u,\psi)+\Re \int_0^T \langle \chi|Z(u,\psi)\rangle \text{d}t, \label{lagrangian}
\end{equation}
where $Z(u,\psi)$ is the constraint given by
\begin{equation}
Z(u,\psi) = i  \dot{\psi}-\hat{H}\psi-\beta |\psi|^2 \psi.
\end{equation}
Clearly this constraint is zero for any $\psi(t)$ that also satisfies the GPE equation. The optimality system is found by requiring that all first order variations vanish. This gives us the equations,
\begin{equation}
D_{\delta \chi} \mathcal{L} = D_{\delta \psi} \mathcal{L} = D_{\delta u} \mathcal{L} = 0,
\end{equation}
for all admissible variations. We now calculate these three variations one by one. The variation with respect to $\chi$ gives,
\begin{align}
D_{\delta \chi} \mathcal{L}&=D_{\delta \chi} \Re \int_0^T\langle \chi |i  \dot{\psi}-\hat{H}\psi-\beta |\psi|^2 \psi\rangle\text{d}t \nonumber \\
&=\Re \int_0^T\langle \delta \chi |i  \dot{\psi}-\hat{H}\psi - \beta |\psi|^2 \psi\rangle \text{d}t.
\end{align}
This variation must be zero for all variations $\delta \chi(t)$, which gives Eq. (\ref{optim1}). Next consider the variation with respect to $\psi$. A variation of the constraint gives
\begin{widetext}
\begin{align}
D_{\delta \psi} \Re \int_0^T \langle \chi|Z\rangle \text{d}t &=  D_{\delta \psi} \biggl( \Re\langle \chi|i \psi\rangle \biggl |_0^T - \Re \int_0^T \langle \dot{\chi}|i \psi\rangle \text{d} t\biggl) - \Re \int_0^T \langle \chi|\hat{H}\delta \psi +\beta\psi^2 \delta \psi^* + 2\beta|\psi|^2 \delta \psi \rangle \text{d}t \nonumber \\
&=  \Re \langle i \chi(0)|\delta \psi(0)\rangle -  \Re \langle i \chi(T)|\delta \psi(T)\rangle + \Re \int_0^T \langle i  \dot{\chi}-\hat{H}\chi-2\beta |\psi|^2 \chi - \beta \psi^2 \chi^*|\delta \psi\rangle. \label{psiVar}
\end{align}
\end{widetext} 
First consider $0<t<T$. Since all variations must vanish we find the optimality condition in Eq. (\ref{optim2}). For $t=T$, the cost functional $J(u,\psi)$ also contributes with the term
\begin{align}
D_{\delta \psi}J&=-\frac{1}{2} D_{\delta \psi} \langle \psi_t|\psi(T)\rangle\langle \psi(T)|\psi_t\rangle \nonumber \nonumber \\
&= - \Re\Bigl(\langle \psi(T)|\psi_t\rangle\langle\psi_t|\delta \psi(T)\rangle\Bigr).
\end{align}
Combining this with Eq. (\ref{psiVar}) gives
\begin{align}
0&=-\Re\biggl[\Bigl( \langle i \chi(T)|+\langle \psi(T)|\psi_t\rangle\langle\psi_t|\Bigr)|\delta \psi(T)\rangle\biggr] \nonumber \\
&+\Re\langle i \chi(0)|\delta \psi(0)\rangle.
\end{align}
The first term in this equation gives us the boundary condition for the Lagrange multiplier in Eq. (\ref{boundary2}). There is also the initial condition that $\psi(0)=\psi_0$ [Eq. (\ref{boundary1})], which gives $\delta \psi(0)=0$ implying that the second term in the equation above is zero. Finally, consider the variation with respect to the control \textit{u}. First, consider the variation with respect to the cost functional,
\begin{align}
D_{\delta u} J&= \gamma(\dot{u}(T)\delta u(T) - \dot{u}(0)\delta u(0)) \label{CostBoundary}\\
&- \gamma \int_0^T \ddot{u} \delta u \text{d}t
\end{align}
The terms at the boundary vanish, since the control must be fixed at the boundary (Eq. (\ref{boundary3})). With this expression, we find the variation of the Lagrangian to be
\begin{equation}
D_{\delta u} \mathcal{L} = -\int_0^T \biggl(\Re \biggl \langle \chi\biggl|\diff{\hat{H}}{u}\biggr|\psi\biggr\rangle + \gamma \ddot{u}\biggr)\delta u\text{d}t. \label{Uvariation}
\end{equation}
Since this must vanish for all variations, we find the last optimality condition in Eq. (\ref{boundary3}).

\subsection{Calculation of GRAPE Gradients} \label{sec:appGRAPE}
In the main text we introduce the gradients for \textsc{grape}. Here we give a brief derivation of these equations. Our considerations follow Refs. \cite{mennemann2015optimal,von2008computational}. In numerical simulations, we solve the GPE for a given \textit{u}, so it is more natural to consider the reduced cost functional 
$\hat{J}(u,\psi)=J(u,\psi_u)$ where $\psi_u$ is the solution to the GPE for a given \textit{u}. Note that $\hat{J}(u)=\mathcal{L}(\psi_u,u,\chi)$ since $Z(u,\psi_u)=0$. This is correct for any $\chi$, hence $\chi$ is a free variable that we can choose conveniently. In order to find the gradient for \textsc{grape} we need the variation of the reduced cost functional
\begin{equation}
D_{\delta u}\hat{J}(u)=D_{\delta u}\mathcal{L}(\psi,u,\chi)+D_{\delta \psi_u} \mathcal{L}(\psi,u,\chi),
\end{equation}
where we have used the total derivative. The derivative $D_{\delta \psi_u} \mathcal{L}$ is the induced variation in $\psi_u$ from the variation of \textit{u}. This extra term appears since $\psi$ depends implicitly on \textit{u} through the GPE equation. This was not the case when discussing the Lagrangian [Eq. \ref{lagrangian}], since $\psi$ was taken to be a free variable and the GPE is a constraint. Notice that performing this variation is formally the same as the $D_{\delta \psi}$, just with the induced variation instead. Hence, we would get the same equations as above for $D_{\delta \psi} \mathcal{L}$ just with $\delta \psi$ replaced with $\delta \psi_u$. Specifically, we find Eq. (\ref{psiVar}) again for the induced variation. Recall that $\chi$ is now a free variable that we can select. If we pick $\chi$ to satisfy Eq. (\ref{optim2}), then the induced variation $D_{\delta \psi_u}$ will vanish, leading to the conclusion that
\begin{align}
   D_{\delta u}\hat{J}(u)&=D_{\delta u}\mathcal{L}(\psi,u,\chi) \nonumber \\
   &= -\int_0^T\biggl( \Re \biggl\langle \chi\biggl| \diff{\hat{H}}{u}\biggr|\psi\biggr\rangle + \gamma \ddot{u}\biggr) \delta u \text{d}t, \label{costVariationU}
\end{align}
where we have used the result from Eq. (\ref{Uvariation}). Recall from the discussion in the main text that the \textsc{grape} gradient is defined as the unique element such that $(\nabla \hat{J},\delta u)_X = D_{\delta u}\hat{J}$. The gradient depends on the choice of the inner product. The typical choice of the $L^2$ inner product gives
\begin{align}
(\nabla \hat{J}, \delta u)_{L^2} &= \int_0^T \nabla \hat{J} \delta u\text{d}t \nonumber \\
&=-\int_0^T\biggl( \Re \biggl\langle \chi\biggl| \diff{\hat{H}}{u}\biggr|\psi\biggr\rangle + \gamma \ddot{u}\biggr) \delta u \text{d}t. 
\end{align}
From this equation we can immediately recognize that
\begin{align}
\nabla \hat{J} = -\Re \biggl\langle \chi\biggl| \diff{\hat{H}}{u}\biggr|\psi\biggr\rangle - \gamma \ddot{u}. \quad \text{for } L^2.
\end{align}
Note that there is no reason for this expression to vanish at the boundaries ($t=0$ and $t=T$). As discussed in the main text failing, to satisfy the boundary conditions for the control [Eq. (\ref{boundary3})] can cause instabilities in the algorithm. It turns out that a more suitable choice is the $H^1$ inner product. Here the gradient is
\begin{align}
(\nabla \hat{J},\delta u)_{H^1}&=\delta u \dif{}{t}\nabla \hat{J} \biggl|_0^T- \int_0^T \delta u \dif{^2}{t^2} \nabla \hat{J} \text{d}t \label{H1eqStep} \\
&= -\int_0^T \biggl( \Re \biggl\langle \chi\biggl| \diff{\hat{H}}{u}\biggr|\psi\biggr\rangle + \gamma \ddot{u}\biggr) \delta u \text{d}t.
\end{align} 
The first term in Eq. (\ref{H1eqStep}) vanishes since $\delta u(0)=\delta u(T)=0$. From this expression we can directly read off the $H^1$ gradient,
\begin{equation}
\dif{^2}{t^2} [\nabla \hat{J}(u)] = \gamma \ddot{u}+\Re\biggl\langle \chi\biggl|\diff{\hat{H}}{u} \biggr| \psi \biggr\rangle \quad \text{for } H^1.
\end{equation}
which is the result given in Eq. (\ref{H1Grad}). This is a Poisson equation, so the Dirichlet boundary condition that the gradients vanish at the boundaries can be chosen. This is the motivation for using the $H^1$ gradient over the $L^2$ gradient.

As discussed in the main text, it is necessary in condensate driving to take the finite bandwidth of the electronics into account. This effect distorts the control $u(t)$ into $v(t)$, which enters into the GPE equation. This can be modeled using a filter function $h(t)$,
\begin{equation}
v(t) = (h*u)(t) = \int_0^t h(\tau) u(t-\tau)\text{d}\tau,
\end{equation}  
where $v(t)$ is the distorted control. Again, we can calculate the variation. This can be done with the chain rule for variations,
\begin{align}
D_{\delta u} \hat{J}(v)&=D_{D_{\delta u}v} \hat{J}(v) \nonumber \\
&=-\int_0^T \biggl( \Re \biggl\langle \chi\biggl| \diff{\hat{H}}{v}\biggr|\psi\biggr\rangle + \gamma \ddot{v}\biggr) (h*\delta u)(t)\text{d}t. \nonumber \\
&+\gamma \dot{v}(T) (h*\delta u)(T) \label{varFilter}
\end{align}
The first term is directly found from Eq. (\ref{Uvariation}) where the expression inside the bracket is evaluated along the distorted control $v(t)$. Note that when calculating the variation of the regularization with respect to \textit{u}, two boundary terms were zero due to the boundary conditions for $\delta u$. However, the first of these terms is not zero when the filter is included. This gives the second term in the equation above. This equation can be rewritten as
\begin{align}
&D_{\delta u} \hat{J}(v)=\int_0^T \delta u(t) \biggl[\gamma \dot{v}(T) h(T-t) \nonumber \\
&-\int_0^T\biggl(\Re \biggl\langle \chi\biggl| \diff{\hat{H}}{v}\biggr|\psi\biggr\rangle + \gamma \ddot{v}\biggr)h(t'-t)\Theta(t'-t)\text{d}t'\biggl]\text{d}t,
\end{align}
where $\Theta(t'-t)$ is the Heaviside-step function. From this expression, we can identify the gradients for both $L^2$ and $H^1$. A similar expression for the filter gradient was reported in Ref. \cite{jager2013optimal}.  With the same arguments as in Eq. (\ref{H1eqStep}), the $H^1$ gradient is given as
\begin{align}
\dif{^2}{t^2} [\nabla \hat{J}(u)] &= -\gamma \dot{v}(T) h(T-t) \nonumber \\
&+\int_t^T \biggl(\Re \biggl\langle \chi\biggl| \diff{\hat{H}}{v}\biggr|\psi\biggr\rangle + \gamma \ddot{v}\biggr)h(t'-t)\text{d}t'
\end{align}
This is the expression given in the main text.

\subsection{Calculation of GROUP Gradients} \label{sec:appGROUPgrad}
Here we give a brief deviation of the \textsc{group} gradient expression presented in the main text. As discussed in the main text in \textsc{group} the control in expressed as the linear combination
\begin{equation}
u(t) = u_0(t) + S(t) \sum_{n=1}^M c_n f_n(t),
\end{equation}
where the $f_n(t)$ are smooth functions. Here the optimization is over the expansion coefficients $c_n$'s. The partial derivative with respect to one of these coefficients can be found using the chain rule for variational derivatives
\begin{align}
\diff{\hat{J}(\mathbf{c})}{c_n}&=D_{S(t)f_n(t)}\hat{J}(u) \nonumber \\
&=-\int_0^T \biggl(\Re \biggl\langle \chi\biggl|\diff{\hat{H}}{u}\biggr|\psi\biggr\rangle + \gamma \ddot{u} \biggr)S(t)f_n(t)\text{d}t,
\end{align}
where we have simply reused the result from Eq. (\ref{Uvariation}) and replaced $\delta u(t)$ with $S(t)f_n(t)$. There is no contribution from the boundary terms in the regularization since $S(0)=S(T)=0$. The chain-rule also allows us to take the filter function into account since we can replace $\delta u(t)$ with $S(t)f_n(t)$ in Eq. (\ref{varFilter}),
\begin{align}
\diff{\hat{J}(v)}{c_n}&=-\int_0^T \biggl(\Re\biggl\langle \chi\biggl|\diff{\hat{H}}{v}\biggr|\psi\biggr\rangle + \gamma \ddot{v} \biggr)(h*Sf_n)(t)\text{d}t \nonumber \\
&+\gamma \dot{v}(T) (h*Sf_n)(T)
\end{align}
The expression inside the bracket is evaluated along the distorted control $v(t)$.

\subsection{GOAT} \label{sec:appGOAT}
In Refs. \cite{machnes2015gradient,de2011second} it was proposed to calculate the gradient in a reduced-basis $\partial \hat J(\mathbf{c})/\partial c_n$ using another method named gradient optimization of analytic controls (\textsc{goat}), which we briefly discuss here. Ignoring the regularization, in Eq. (\ref{costFun}) the derivative of the cost is
\begin{equation}
\diff{\hat J}{c_n} = - \Re \biggl \langle \psi_t \biggr| \diff{\psi}{c_i}(T)\biggr\rangle.
\end{equation}
In Ref. \cite{machnes2015gradient} the derivative of the state is directly computed from the equations of motion, which gives
\begin{equation}
i \partial_t
\begin{pmatrix}
\psi \\
\partial_{c_n} \psi
\end{pmatrix} 
=
\begin{pmatrix}
H+\beta |\psi|^2 & 0 \\
\partial_{c_n} H +\beta \psi \partial_{c_n} \psi^* & H + 2 \beta|\psi|^2
\end{pmatrix}
\begin{pmatrix}
\psi \\
\partial_{c_n} \psi
\end{pmatrix}. \label{goatEquations}
\end{equation}
This is the straightforward extension of the method from Ref. \cite{machnes2015gradient} to the nonlinear dynamics of the GPE and for $\beta=0$ we obtain the original result. In order to compute the derivative in a basis with size $M=60$, it would then be necessary to compute $\partial_{c_1} \psi$,$\partial_{c_2} \psi$,...,$\partial_{c_{60}}\psi$ using the relation above. As we discussed in the main text $M=60$, is numerically found to be the optimal basis size. This can be done by constructing one huge matrix of size $((M+1)d)\times(2d)$, where \textit{d} is the number of grid points used when discretizing the GPE. An alternative is to calculate 60 independent solutions of Eq. (\ref{goatEquations}), which could be done in parallel. Compared with the expression in Eq. (\ref{groupGRAD}) that only requires integrating two equations of motion, calculating the gradient using \textsc{goat} is much more demanding. Nevertheless, the \textsc{goat} approach potentially offers a higher numerical accuracy in the gradient, since the accuracy of Eq. (\ref{groupGRAD}) is limited by the error from discretizing the time evolution. This is an advantage when optimizing for very low errors ($1-F\leq 10^{-9}$) as needed for some error-correction protocols in quantum computing \cite{machnes2015gradient}. Very low infidelities below $10^{-4}$ are not required in the condensate driving control problem, so calculating the derivative using the more expensive \textsc{goat} method would most likely not be an advantage. A fair comparison between these two methods for calculating the gradient would require studying a wider class of problems and discussing other alternatives for calculating high-accuracy gradients, which is beyond the scope of this paper.

\subsection{Krotov's Method} \label{sec:appKrotov}
\begin{figure}[t]
\includegraphics[width=\columnwidth]{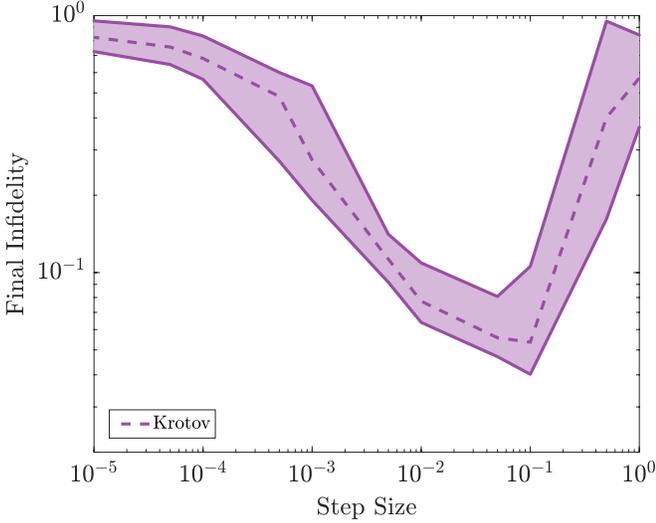}
\caption{The final infidelity after 2500 evaluations for Krotov as a function of the step size ($\alpha$) used in Eq. (\ref{krotovUpdate}). The dotted line and shaded area show the median, 25\% and 75\% quartiles for the 100 initial controls.}
\label{fig:krotovStepSize}
\end{figure}
As discussed in the main text Krotov's method is an alternative to the Lagrange multiplier approach for deriving optimal control algorithms \cite{konnov1999global,reich2012monotonically}. Here we briefly present Krotov's method as described in Refs. \cite{reich2012monotonically,sklarz2002loading,eitan2011optimal}. We do not include the regularization term in Krotov so the cost functional becomes $J_T = (1-|\langle \psi_t|\psi(T)\rangle|^2)/2 $. In Krotov's method, the cost functional is rewritten so the GPE appears explicitly. This is done by adding the vanishing quantity $0 = \phi(T)-\phi(0)-\int_0^T \dif{\phi}{t} \text{d}t$. Here $\phi$ is some arbitrary field, which we can select freely. In order to present the derivation more clearly we write the variational derivatives differently here and adopt the standard notation that $\dot{\psi}=f(\psi,u)=-i\hat{H}_{\text{NL}}\psi$ where $\hat{H}_{\text{NL}}$ is the GPE Hamiltonian from Eq. (\ref{GPE}). This allows the cost functional to be rewritten as,
\begin{equation}
L = G(\psi(T))-\phi(0) - \int_0^T R(\psi,u) \text{d}t
\end{equation}
where
\begin{gather}
G(\psi(T))=J_T + \phi(T), \label{defnG} \\
R(\psi,u)=\diff{\phi}{t}+\vardiff{\phi}{\psi}f+f^*\vardiff{\phi}{\psi^*},
\end{gather}
From the arguments above and the definition of \textit{R} it is seen that $L=J$ as discussed in Ref. \cite{sklarz2002loading}. So minimizing \textit{J} is equivalent to minimizing \textit{L}. In Krotov's method, it is directly required that the cost decreases at each iteration $J^{(i+1)}\leq J^{(i)}$, which is equivalent to $0\leq L^{(i)}-L^{(i+1)} = \Delta_1 + \Delta_2 + \Delta_3$ where,
\begin{align}
\Delta_1 &= G(\psi^{(i)}(T))- G(\psi^{(i+1)}(T)), \\
\Delta_2 &= \int_0^T R(\psi^{(i+1)},u^{(i+1)})-R(\psi^{(i+1)},u^{(i)})\text{d}t,\\
\Delta_3 &= \int_0^T R(\psi^{(i+1)},u^{(i)})-R(\psi^{(i)},u^{(i)}) \text{d}t.
\end{align}
A sufficient condition for a decrease in the cost is that the $\Delta$'s are positive. Central in Krotov's method is the unintuitive notion that \textit{L} is maximized with respect to the states at the current iteration. This implies that any change in the states caused by selecting a new value of the control for the next iteration would decrease the value of \textit{L} \cite{sklarz2002loading,reich2012monotonically}. If \textit{L} is maximal with respect to the states, then the first order derivatives with respect to $\psi^*$ must vanish. These derivatives are,
\begin{align}
\vardiff{R}{\psi^*}&=\biggl( \diff{}{t}+f\vardiff{}{\psi}+f^*\vardiff{}{\psi^*} \biggr) \vardiff{\phi}{\psi^*}+\vardiff{\phi}{\psi}\vardiff{f}{\psi^*}+\vardiff{f^*}{\psi^*}\vardiff{\phi}{\psi^*} \nonumber \\
&=\dif{}{t}\vardiff{\phi}{\psi^*}+\vardiff{\phi}{\psi}\vardiff{f}{\psi^*}+\vardiff{f^*}{\psi^*}\vardiff{\phi}{\psi^*},
\end{align}
and
\begin{align}
\vardiff{G(\psi(T))}{\psi^*} = \vardiff{J_T}{\psi^*}+\vardiff{\phi(T)}{\psi^*}. \label{Gcond}
\end{align}
From the discussion above we require that $\delta R/\delta \psi^*|_{\psi^{(i)},u^{(i)}}=0$ and $\delta G/\delta \psi^*|_{\psi^{(i)},u^{(i)}}=0$. This condition gives for \textit{R} that
\begin{align}
\dif{}{t}\vardiff{\phi}{\psi^*}\biggl|_{\psi^{(i)},u^{(i)}} = \biggl(&i \psi \vardiff{\hat{H}_{\text{NL}}}{\psi^*}\vardiff{\phi}{\psi}-i\psi^*\vardiff{\hat{H}_{\text{NL}}}{\psi^*}\vardiff{\phi}{\psi^*} \nonumber \\ 
&-i\hat{H}_{\text{NL}} \vardiff{\phi}{\psi^*}\biggr)\biggl|_{\psi^{(i)},u^{(i)}}. \label{Rcond}
\end{align}
The requirement that \textit{L} is maximal with respect to $\psi$ also puts requirements on the second-order derivatives of \textit{R} and \textit{G}. A good ansatz for $\phi$ is a second-order expansion in the states
\begin{equation}
\phi = \frac{1}{2} \biggl( \langle \xi|\psi\rangle + \langle \psi|\xi\rangle\biggr) + \frac{1}{4} \langle \Delta \psi| \sigma(t) | \Delta \psi\rangle,
\end{equation}
where $\xi$ are some expansion coefficients, $\sigma(t)$ is an arbitrary time-dependent function, and $\Delta \psi = \psi - \psi^{(i)}$ is the difference from the next iteration to the current. If this expression is inserted into Eqs. (\ref{Gcond}) and (\ref{Rcond}), one obtains
\begin{gather}
i\dot{\xi}^{(i)}=(\hat{H}(u^{(i)})+2\beta|\psi^{(i)}|^2)\xi^{(i)} - (\psi^{(i)})^2 \beta \xi^{(i)*} \\
\xi^{(i)}(T)=-|\psi_t\rangle\langle\psi_t|\psi^{(i)}(T)\rangle. \label{optimKrotov2}
\end{gather}
Incidentally, these equations are the same as those for $\chi$ given in Eqs. (\ref{optim2}) and (\ref{boundary2}) with an extra phase factor, so we have $\xi=i\chi$. We now discuss the additional conditions that ensure that each $\Delta$ is positive.

The boundary condition for $\xi$ [Eq. (\ref{optimKrotov2})] can be combined with the definition of \textit{G} Eq. (\ref{defnG}), which gives
\begin{align}
G(\psi(T))&=\frac{1}{2}\Bigl(1-\langle \psi(T)|\hat{P}|\psi(T)\rangle \nonumber \\
&+\langle\psi^{(i)}(T)|\hat{P}|\psi(T)\rangle+\langle\psi(T)|\hat{P}|\psi^{(i)}(T)\rangle\Bigr)
\end{align}
where $\hat{P}=|\psi_t\rangle\langle\psi_t|$ is the projection operator for the target state. From this expression $\Delta_1$ can be directly rewritten as $2\Delta_1 = \langle \Delta \psi |\hat{P}|\Delta \psi\rangle$, which is always non-negative due to the positivity of $\hat{P}$.

Generally, it is more difficult to ensure that $\Delta_3$ is positive. Additional conditions on the second-order derivatives with respect to the states on \textit{R} are also required, e.g., $\delta^2 R / \delta \psi \delta \psi^*|_{\psi^{(i)}}>0$. This can be ensured with a proper choice of the $\sigma(t)$ function. A number of different strategies for choosing $\sigma(t)$ have been discussed in the literature \cite{sklarz2002loading,reich2012monotonically,jager2014optimal}. For the moderate values of $\beta$ discussed here, a good strategy is simply to select $\sigma(t)=0$ and forfeit the strict guarantee that $\Delta_3$ is positive \cite{jager2014optimal}. Note that if $\beta=0$, then \textit{R} is independent of $\psi$ and $\Delta_3=0$. If $\Delta_3=0$, we can just pick $\sigma(t)=0$ and the algorithm is guaranteed to decrease the cost at every iteration.

Finally, we discuss how to ensure $\Delta_2$ is positive by a proper update for the control. Ideally, we seek a control such that the derivative vanishes ($\partial R/\partial u|_{u^{(i+1)},\psi^{(i+1)}}=0$). This derivative is,
\begin{align}
\diff{R}{u}\biggl|_{u^{(i+1)},\psi^{(i+1)}}&=\Im\biggl[\biggl\langle \xi\biggl|\diff{\hat{H}}{u}\biggr|\psi\biggr\rangle \nonumber \\
&+ \frac{\sigma(t)}{4} \biggl\langle \Delta \psi \biggl|\diff{\hat{H}}{u}\biggr|\psi\biggr\rangle\biggl]\biggl|_{u^{(i+1)},\psi^{(i+1)}}
\end{align}
It is difficult to ensure that this is zero \cite{reich2012monotonically}. Instead of choosing the optimal control, we take a small step in the direction of the gradient,
\begin{equation}
u^{(i+1)}=u^{(i)}+\alpha S(t) \diff{R}{u}\biggl|_{\psi^{(i+1)},u^{(i)}}, \label{krotovAppendix}
\end{equation}
where $\alpha>0$ and $S(t)$ is the shape function that vanishes for $t=0$ and $t=T$ so the boundary conditions on the control can be satisfied [Eq. (\ref{boundary3})].
If $R(\psi^{(i+1)},u^{(i+1)})$ in $\Delta_2$ is Taylor expanded around the last control, we find
\begin{align}
\Delta_2 &\approx \int_0^T \diff{R}{u}\biggl|_{\psi^{(i+1)},u^{(i)}}(u^{(i+1)}-u^{(i)})\text{d}t \nonumber \\
&=\int_0^T \alpha S(t) \biggl(\diff{R}{u}\biggl|_{\psi^{(i+1)},u^{(i)}}\biggr)^2 \text{d}t \geq 0
\end{align}
These arguments show that if the control is chosen as in Eq. (\ref{krotovAppendix}), then all three $\Delta's$ are positive and the cost will decrease. If Eq. (\ref{krotovAppendix}) is rewritten with $\chi$ then Eq. (\ref{krotovUpdate}) given in the main text is found. 

A proper value of the step size $\alpha$ must also be selected in order to ensure a fast convergence. The performance of the Krotov algorithm in condensate driving is shown for different step sizes in Fig. \ref{fig:krotovStepSize}. If the step size is too small, the algorithm converges slowly due to the small steps in the update. On the other hand, if the steps are too large, then there is no longer a guarantee that $\Delta_3$ is positive and the update might not decrease the cost. The optimal step size is around 0.1. 
\end{document}